\documentclass[a4paper,11pt]{article}%
\usepackage{amssymb}
\usepackage{graphicx}
\usepackage{amsbsy}
\usepackage{amsmath}
\usepackage{cite}
\usepackage{slashed}
\usepackage{amsfonts}
\usepackage{hyperref}
\usepackage{comment}
\usepackage[utf8]{inputenc}

\usepackage{multicol}
\usepackage{xcolor}
\definecolor{verdes}{cmyk}{0.92,0,0.59,0.4}
\definecolor{Grn}{rgb}{0.1,0.5,0.2}
\definecolor{Blu}{rgb}{0.,0.,1.}
\definecolor{Red}{rgb}{0.7,0.1,0.1}
\definecolor{SE}{rgb}{0.5,0,0.4}

\newcommand{\Red}[1]{{\color{Red}{#1}}}

\usepackage{hyperref} 
\hypersetup{
     colorlinks = true,
     linkcolor = blue,
     citecolor = verdes,
     anchorcolor = blue,
     filecolor = blue,
     urlcolor = verdes,
     bookmarks=true,
     linktocpage =true,
     }

\newcommand{\ket}[1]{\left| #1 \right>} 
\newcommand{\bra}[1]{\left< #1 \right|} 
\newcommand{\braket}[2]{\left< #1 \vphantom{#2} \right|
 \left. #2 \vphantom{#1} \right>} 

\usepackage{dsfont}

\def\logm#1{\log\frac{#1}{\mu^2}}

\newcommand{\tr}{\text{tr}}
\newcommand{\Integral}{\mathcal{I}}

\usepackage{tabularx}

\newcommand\T{\rule{0pt}{3.2ex}}
\newcommand\B{\rule[-2.1ex]{0pt}{0pt}}

\usepackage[listings]{tcolorbox}

\usepackage{mathtools}

\usepackage{dsfont}
\usepackage{feynmp}
\usepackage{longtable}
\usepackage{multirow}

\usepackage{hhline}
\usepackage{rotating}
\usepackage{array}
\usepackage{float}
\usepackage{afterpage}

  {}

\allowdisplaybreaks

\begin{document}
\thispagestyle{empty} \setcounter{page}{0} \begin{flushright}
December 2021\\
\end{flushright}

\vskip          4.1 true cm

\begin{center}
{\huge Axion Effective Action}\\[1.9cm]

\textsc{J\'er\'emie Quevillon}$^{1}$, \textsc{Christopher Smith}$^{2}$ \textsc{\ and Pham Ngoc Hoa Vuong}$^{3}$%
\vspace{0.5cm}\\[9pt]\smallskip{\small \textsl{\textit{Laboratoire de
Physique Subatomique et de Cosmologie, }}}\linebreak%
{\small \textsl{\textit{Universit\'{e} Grenoble-Alpes, CNRS/IN2P3, Grenoble
INP, 38000 Grenoble, France}.}} \\[1.9cm]\textbf{Abstract}\smallskip
\end{center}

\begin{quote}

In this paper, we discuss the construction of Effective Field Theories (EFTs) in which a chiral fermion, charged under both gauge and global symmetries, is integrated out. Inspired by typical axion models, these symmetries can be spontaneously broken, and the global ones might also be anomalous. In this context, particular emphasis is laid on the derivative couplings of the Goldstone bosons to the fermions, as these lead to severe divergences and ambiguities when building the EFT. We show how to precisely solve these difficulties within the path integral formalism, by adapting the anomalous Ward identities to the EFT context. Our results are very generic, and when applied to axion models, they reproduce the non-intuitive couplings between the massive SM gauge fields and the axion. Altogether, this provides an efficient formalism, paving the way for a systematic and consistent methodology to build entire EFTs involving anomalous symmetries, as required for axion or ALP searches.

\let\thefootnote\relax\footnotetext{$^{1}$jeremie.quevillon@lpsc.in2p3.fr}\footnotetext{$^{2}$chsmith@lpsc.in2p3.fr}\footnotetext{$^{3}$ vuong@lpsc.in2p3.fr}
\end{quote}

\newpage

\setcounter{tocdepth}{3}
\tableofcontents

\newpage
\section{Introduction}

The Peccei-Quinn (PQ) mechanism~\cite{Peccei:1977hh,Peccei:1977ur} is probably the best solution to the strong CP problem of the Standard Model (SM)~\cite{Abel:2020gbr}.  This solution predicts a new Goldstone boson, the so-called axion~\cite{Weinberg:1977ma,Wilczek:1977pj}, which is hunted by many experiments. Recent constrains from astrophysics~\cite{Grifols:1996id,Brockway:1996yr} and particle physics~\cite{Kim:1986ax,Mimasu:2014nea} imply that the breaking scale of PQ symmetry, $f_a$, is much larger than the electroweak scale.  Due to this large-scale separation,  Effective Field Theory (EFT) is a well-suited framework to describe the interactions between the axion and other light particles (usually from the SM). 

In previous works, some of us have shown~\cite{Quevillon:2019zrd, Quevillon:2020hmx,Quevillon:2020aij} that axion models exhibit intrinsic ambiguities in their formulation, and this has a dramatic impact on the coupling of axions to massive gauge fields. One of the main conclusion of Ref.~\cite{Quevillon:2019zrd} states as follows: when axion models are specified in a representation in which the axion has only derivative couplings to SM chiral fermions, such as in DFSZ-like axion models~\cite{Dine:1981rt,Zhitnitsky:1980tq}~\footnote{KSVZ-like models~\cite{Kim:1979if,Shifman:1979if} involve vector-like fermions, whose masses are decoupled from the spontaneous electroweak symmetry breaking, and the discussion is much simpler.}, some chiral reparametrisation of the fermionic fields are implicit and lie at the root of the so-called anomalous axion couplings to gauge field strengths. For vector gauge interactions, it is well known that derivative couplings to fermions decouple faster than local anomalous operators, which thus capture the whole axion to gauge boson couplings. By contrast, derivative interactions do not systematically decouple for chiral gauge interactions, ultimately because the gauge symmetry is necessarily spontaneously broken when the chiral fermions get their masses. Importantly, non-decoupling contributions from derivative interactions can arise from the usual axial coupling to fermions, but also from the vector one. Both can be anomalous in the presence of chiral gauge interactions. In practice, keeping track of these non-decoupling effects is crucial to get consistent, parametrisation-independent couplings of the axion to gauge bosons. Only with them, one can match the results obtained using a linear representation of the complex Peccei-Quinn scalar field, in which the axion has pseudoscalar couplings to chiral fermions, and no anomaly-related ambiguities ever arise.

On a more technical side, these results have been derived by appropriately computing anomalous triangle diagrams regularised using Weinberg's method \cite{Weinberg:1996kr,Bilal:2008qx} which allow to parametrise the initial ambiguity inherent to the momentum rooting in the amplitude integrals. This rigorous treatment allows to obtain generalised Ward identities, in which one can tune which current is anomalous, or not, which is physically mandatory and not guaranteed in a more naive computation. The Ref.~\cite{Bonnefoy:2020gyh} reaches similar conclusions from a more anomaly matching EFT-oriented point of view, which brings interesting insights to axion couplings to chiral gauge fields.

In this work, our goal is not only to add up on the understanding and construction of low energy axion EFTs, but also more generally on the possible interplays or entanglements between spontaneous and anomalous symmetry breaking that can arise when chiral fermions are integrated out. Further, our goal is to perform this analysis exclusively in a functional context, by building the low-energy EFT following a step-by-step integration of the chiral fermion fields, without recourse to triangle Feynman diagrams or Ward identities, and take advantage of the elegant and convenient techniques developed recently to integrate out heavy fermionic fields \cite{Henning:2014wua,Drozd:2015rsp,Fuentes-Martin:2016uol,Zhang:2016pja,Ellis:2016enq,Henning:2016lyp,Ellis:2017jns,Kramer:2019fwz, Ellis:2020ivx,Angelescu:2020yzf,Cohen:2020fcu}. The only ingredients will thus be dimensionally-regulated functional traces, and the order-by-order invariance of the EFT operators under gauge transformations, when the appropriate would-be-Goldstone bosons are accounted for. Ultimately, the same non-decoupling of derivative interactions will be observed, in the sense that the EFT built from them will start with dimension-five operators.

The main novelties of our approach are the following:  within the path integral formalism for one-loop matching, we show how to consistently integrate out heavy chiral fermions, which are charged under both the local and global symmetries. Focusing on the Goldstone-gauge bosons couplings,  we show how to deal with $\gamma^5$ within dimensional regularisation to properly keep track of the ambiguities arising in the one-loop effective action.   In the functional matching,  we show that the gauge invariant combinations of the EFT operators can be used to fix these ambiguities.  Hence, for the first time, with the functional method for one-loop matching, we can fully control which symmetry currents are anomalous and which ones are anomaly free.  Therefore, the Wilson coefficients can be obtained correctly.  We then derive the universal formula that captures all EFT couplings of Goldstone bosons with gauge bosons (both massive and massless) consistently and generically.  Our results can be easily applied to various axion UV models.

The plan of the manuscript is the following: In section \ref{toymodel}, we integrate out a chiral fermion from a toy model to obtain a gauge and Goldstone boson EFT. This section will mainly concentrate on the physical interpretations so the reader can understand the logic behind the construction without entering into the details of the calculation. The crucial point of this section is to show how ambiguous coefficients can be fixed by enforcing the Ward identities, which are now written in terms of gauge invariant combinations of the EFT operators.  For the reader who is familiar to this topic,  one can skip this section and go directly to the following section.

In section \ref{UOLEA}, we will detail how to evaluate the one-loop effective action from the path integral functional approach and how we deal with the ambiguities originating from the QFT anomalies. The main outcome of this section is Eq.~\eqref{masterf}, a master formula which can be easily used to obtain effective couplings between gauge fields and Goldstone bosons, and which encapsulates the subtleties occurring when dealing with anomalous global symmetries in a chiral gauge theory. 

In section \ref{section: application}, we apply this master formula to various models starting with a simple chiral toy model with an additional global $U(1)$ symmetry. We then explicitly apply our results to axion models and recover, for instance, the non-intuitive axion couplings involving massive gauge fields in DFSZ-like models. We conclude in section \ref{Ccl} while additional computational details regarding master integrals can be found in appendix \ref{Appendix:master_integrals}.

\section{EFTs with spontaneously and anomalously broken symmetries}
\label{toymodel}
Readers familiar to the topic of anomalous symmetries in the EFT may directly jump to the following sections. Our goal here is to introduce the formalism using a simplified setting. More precisely, our goal is to integrate out fermions that can be charged under both global and local symmetries. Further, those fermions will not be assumed vector-like: their left- and right-handed components need not have the same charges under these symmetries. This generates two complications. First, obviously, such fermions can only acquire a mass, and thus be integrated out, when the chiral components of the symmetries are spontaneously broken. Second, the classical symmetries cannot all survive quantisation, and there must be some anomalies. These two effects are entangled, and further, they induce some freedom in how the fermionic part of the UV Lagrangian is to be parametrised. So, before any attempt at integrating out the fermions, it is necessary to fix this freedom. As we will discuss in this section, from a functional point of view, one parametrisation emerges as the most natural, but requires a specific treatment of anomalous effects and derivative interactions. 

\subsection{EFTs and gauge invariance}

We start from a generic UV Lagrangian exhibiting some set of local symmetries and involving fermionic degrees of freedom. Typically, the fermionic part of the Lagrangian is of the form, including for simplicity only one axial and one vector gauge field,  
\begin{align}
    \mathcal{L}_{\rm UV}^{\rm fermion} &= \bar{\Psi} \big( i\partial_{\mu}\gamma^{\mu} + g_{_V}V_{\mu}\gamma^{\mu} - g_{_A}A_{\mu}\gamma^{\mu}\gamma^5 \big) \Psi
    \, .
    \label{Lagrangian: UV toy model}
\end{align}
Let us first consider abelian gauge symmetries for simplicity~\footnote{We will discuss later the peculiarities arising in the non abelian case.}. At the classical level, this theory is invariant under $U(1)_V$ and $U(1)_A$ gauge transformations, which we define as:
\begin{align}
    U(1)_V & : ~ V_{\mu} \rightarrow V_{\mu} + \dfrac{1}{g_{_V}} \partial_{\mu}\theta_{_V} ~ , ~ \Psi \rightarrow \exp( i\theta_{_V}) \Psi
    \, , \\
    U(1)_A & : ~ A_{\mu} \rightarrow A_{\mu} + \dfrac{1}{g_{_A}} \partial_{\mu}\theta_{_A} ~ , ~ \Psi \rightarrow \exp(- i\theta_{_A}\gamma^5 ) \Psi 
    \, .
    \label{GaugeTF1}
\end{align}
Our goal is to integrate out the fermion to get the tower of effective
interactions by performing an inverse mass expansion~\footnote{More precisely we will use convenient Covariant Derivative Expansion (CDE) techniques.}. This obviously means that the
fermion to be integrated out should be massive, which forces the axial gauge
symmetry to be spontaneously broken. To be able to consistently account for
this, let us include the complex scalar field $\phi_A$ which, by acquiring a vacuum expectation value $v$, will spontaneously break the axial gauge symmetry, 
\begin{align}
\mathcal{L}_{\rm UV}^{\rm fermion} = \bar{\Psi} \big( i\partial_{\mu}\gamma^{\mu} + g_{_V}V_{\mu}\gamma^{\mu} - g_{_A}A_{\mu}\gamma^{\mu}\gamma^5 \big) \Psi -y_{\Psi}\big(\bar{\Psi}_{_L}\phi_A\Psi_{_R} + \text{h.c.} \big) \, ,
    \label{LUVPseudo}
\end{align}
with $y_{\Psi}$ the Yukawa coupling, and two Weyl components $\Psi_{R,L} = P_{R,L} \, \Psi$ with $P_{R,L} = ( 1 \pm \gamma^5 )/2$. 

If one wants to focus on manifest gauge invariance, it is convenient to include the Goldstone boson, $\pi_A$, and adopt an exponential or polar representation for the complex scalar field,
\begin{align}
\phi_{A} = \frac{1}{\sqrt{2}}(v+\sigma_{A}) \exp\bigg[ i\, \dfrac{\pi_A(x)}{v} \bigg]
    \, .
    \label{NLgold}
\end{align}
Indeed, thanks to the exponential parametrisation of the Goldstone boson, this theory is still manifestly gauge invariant provided, together with the transformation of Eq.~(\ref{GaugeTF1}),
\begin{align}
    \pi_A \rightarrow \pi_A + 2 v\theta_{_A}
    \, ,
    \label{GaugeTF2}
\end{align}
while $\sigma_{A}$ is gauge invariant and plays no rôle in that regard. Said differently, with this representation, it is sufficient to keep only the gauge bosons and the Goldstone fields explicitly to construct the EFT, which will involve only these fields in a gauge invariant way.

By contrast, if one insists on manifest renormalisability, the Goldstone boson has to enter linearly, that is, by writing the scalar field acquiring a vacuum expectation value $v$ as linear in all its components,
\begin{align}
\phi_A=\frac{1}{\sqrt{2}} (v+\sigma_{A}+i\pi_{A})
    \, . 
    \label{Lgold}
\end{align}
The $\sigma_{A}$ is no longer gauge invariant since a $U(1)_{A}$ gauge transformation is nothing but a $SO(2)$ rotation for the $(v+\sigma_{A},\pi_{A})$ vector. If one insists on manifest gauge invariance, the difficulty then is that $\sigma_{A}$ should explicitly appear in $\mathcal{L}_{\rm UV}^{\rm fermion}$. Even if in the abelian case, this is quite simple, this would introduce unnecessary model-dependence in the non-abelian case.

So, at the end of the day, it is legitimate in order to be able to consistently account for spontaneous breaking of the axial gauge symmetry, to consider the exponential representation of the Goldstone boson,
\begin{align}
    \mathcal{L}_{\rm UV}^{\rm fermion} = \bar{\Psi} \bigg( i\partial_{\mu}\gamma^{\mu} + g_{_V}V_{\mu}\gamma^{\mu} - g_{_A}A_{\mu}\gamma^{\mu}\gamma^5 - M \exp\bigg[ i\,\frac{\pi_A}{v}\gamma^5\bigg] \bigg) \Psi
    \, , 
    \label{LUVPseudo} 
\end{align}
where $M \equiv y_{\Psi}v/\sqrt{2} $ stands for the mass of the fermion. Yet, at this stage, a Taylor expansion\footnote{For the purpose of evaluating the one-loop effective action using the Covariant Derivative Expansion (CDE), truncating this expansion is perfectly consistent since operators at most linear in a given Goldstone boson will be considered. Issues related to the apparent non-renormalisability of the exponential parametrisation will not affect our developments.} produces the pseudoscalar $\pi_{A}\bar{\Psi}\gamma^5\Psi$ coupling, and is the same as it would be in the linear representation of $\phi_A$. So, the distinction between linear and polar representation may appear quite academic. Yet, the exponential parametrisation offers an alternative route. Instead of a Taylor expansion, there is a well-known exact procedure to recover a linearised Lagrangian that allows to transfer the Goldstone dependence from the Yukawa sector to the gauge sector. Based on the chiral rotation that is given by Eq.~\eqref{GaugeTF1}, it suffices to perform a field-dependent reparametrisation of the fermion fields
\begin{align}
    \Psi \rightarrow \Psi = \exp\bigg[-i\dfrac{\pi_{_A}(x)}{2 v}\gamma^5 \bigg] \Psi
    \, ,
    \label{FermionRepar}
\end{align}
and the Lagrangian in Eq.~\eqref{LUVPseudo} becomes
\begin{align}
    \mathcal{L}_{\rm UV}^{\rm fermion} &= \bar{\Psi}\bigg( i\partial_{\mu}\gamma^{\mu} -M  + g_{_V}V_{\mu}\gamma^{\mu} - \bigg[g_{_A} A_{\mu} - \dfrac{\partial_{\mu}\pi_A(x)}{2 v} \bigg]\gamma^{\mu}\gamma^5 \bigg)\Psi
    \, . 
    \label{LUVDer}
\end{align}
Under this form, $\Psi$ is invariant under the axial gauge transformation $U(1)_A$, so the mass term does not cause any trouble even for a chiral gauge symmetry and could easily be factored out for an EFT mass expansion.
The quadratic operator defined in Eq.~\eqref{LUVDer} has the virtue of being manifestly gauge invariant. The Goldstone boson itself ensures the theory stays invariant when $A_{\mu}\rightarrow A_{\mu} + \frac{1}{g_{_A}}\partial_{\mu}\theta_{_A}$ thanks to $\pi_{A}\rightarrow\pi_{A} + 2v\theta_{_A}$. Evidently, for that to work, one should not get rid of them by moving to the unitary gauge. 

However, as a side effect, the theory is still not manifestly renormalisable since the $\partial_{\mu}\pi_A(x)$ operator is of dimension five. Yet, this form looks particularly well suited for an inverse mass expansion since $M\sim v$. Let us stress, though, that one should not be tempted to conclude that the $\partial_{\mu}\pi_A(x)$ operator is subleading and can be neglected. Such considerations can only be consistently done after the fermion field has been integrated out, and as we will see in details in the following, this operator does contribute in general to the leading terms in the EFT.

\subsection{EFTs and anomalies}

The Lagrangian in Eq.~\eqref{LUVDer} looks promising, but to reach it, we had to reparametrise the fermion field, Eq.~(\ref{FermionRepar}), and there is one crucial caveat for that. The fermion being chiral, this reparametrisation does not leave the path integral fermionic measure invariant. In general, given that $\Psi$ is coupled to gauge fields, the Jacobian, obtained using the singlet anomaly result for chiral fermions~\cite{Bilal:2008qx,Bertlmann:1996xk}, sums up to additional terms in the Lagrangian of the form 
\begin{align}
    \mathcal{L}_{\rm UV} \supset \mathcal{L}_{\rm UV}^{\rm Jac} = \dfrac{1}{8\pi^{2}} \dfrac{\pi_{A}}{2v}  \bigg[g_{_V}^2 F_{_V,\mu\nu} \tilde{F}_{_V}^{\mu\nu} + \dfrac{1}{3} g_{_A}^2 F_{_A,\mu\nu} \tilde{F}_{_A}^{\mu\nu} \bigg]
    \, , 
    \label{LUVJac}
\end{align}
with $F_X^{\mu\nu}=\partial^{\mu}X^{\nu}-\partial^{\nu}X^{\mu}$ the usual field strength tensor applied to the generic gauge field X and $\tilde{F}_X^{\mu\nu}=(1/2)\epsilon_{\mu\nu\rho\sigma}F_X^{\rho\sigma}$ its dual field strength tensor with the suffix indicating if this apply to the vector gauge field (V) or the axial one (A). These terms explicitly break gauge invariance, since they get shifted under $\pi_{A}\rightarrow\pi_{A} + 2v\theta_{_A}$. 

There are two main ways to deal with the anomalous contributions shown in Eq.~\eqref{LUVJac}. If one wants to hold the interactions to be gauged, a first possibility consists in tuning the chiral fermionic content such that the total contribution to the anomaly vanishes (as it happens in the SM). The second possibility is to give up gauge invariance and reconsider the local symmetry as a global symmetry. We clarify in the following these two cases to consider:
\begin{itemize}
\item For gauge interactions that are meant to
exist at the quantum level (then not being anomalous), the fermionic content is supposed to be just right so that the sum of all Jacobian terms sum up to zero. As is well known, this is the prototype of the gauge interactions in the SM, where gauge anomalies cancel out only when all matter fields are summed over. The important point is that the corresponding Goldstone fields are allowed to be moved to and from the mass terms without generating a Jacobian contribution. Indeed, the reparametrisation in Eq.~(\ref{FermionRepar}) must not generate Jacobian terms since a gauge transformation acts like that on fermions, see Eq.~(\ref{GaugeTF1}). In this context, the strict equivalence between the $\bar{\Psi}\big(\partial_{\mu}\pi_{A}\gamma^{\mu}\gamma^{5}\big)\Psi$ and $\bar{\Psi}\big(M\gamma_{5}\pi_{A}/v\big)\Psi$ couplings can be viewed as the transcription of the non-anomalous Ward identity $\partial_{\mu}A^{\mu}=2iMP$ with $A^{\mu}=\bar{\Psi}\gamma^{\mu}\gamma^{5}\Psi$ and $P=\bar{\Psi}\gamma^{5}\Psi$. Indeed, to the divergence of any correlation function of the axial gauge current, $\bra{0} A^{\mu}{...} \ket{0}$, we can associate that with $\partial_{\mu}\pi_{A}$ from
Eq.~(\ref{LUVDer}), which can then be equivalently calculated from
Eq.~(\ref{LUVPseudo}) (after Taylor expanding the exponential term). 
Regarding the vector gauge interactions, the situation is simpler since the mass term is gauge invariant. Imposing $V_{\mu}\rightarrow V_{\mu}+\frac{1}{g_{_V}}\partial_{\mu}\theta_{_V}$ requires the $\partial_{\mu}\theta_{_V}$ piece to cancel out, i.e. any correlation function of the vector gauge currents $V^{\mu}=\bar{\Psi}\gamma^{\mu}\Psi$ satisfies the non-anomalous Ward identity $\partial_{\mu}V^{\mu} = 0$.
\item Some of the gauge interactions may simply be absent if their symmetry is kept global. In that case, one can simply remove the corresponding
$\mathrm{A}_{\mu}$ from the Lagrangian, but keep the Goldstone bosons since
they become independent physical degrees of freedom. These global symmetries may or may not have anomalies, but whenever they do, one should keep track of the Jacobian when passing from pseudoscalar to derivative Goldstone boson couplings to fermions. As explained in Refs.\cite{Quevillon:2019zrd, Quevillon:2020hmx}, one must obtain the same results using either the Lagrangian with pseudoscalar couplings (after Taylor expanding the exponential term in Eq.~\eqref{LUVPseudo}), or that using derivative couplings, Eq.~(\ref{LUVDer}), provided the local
anomalous terms, Eq.~(\ref{LUVJac}), are then also included. Indeed, the point is that derivative couplings do induce anomalous effects that precisely cancel those in the local terms of Eq.~(\ref{LUVJac}). In the inverse mass expansion context, this shows that one must be careful not to perform the limit $M\rightarrow\infty$ too soon, that is, discard the derivative interaction in Eq.~(\ref{LUVDer}) on the basis of its relative $\mathcal{O}(M^{2})$ suppression with respect to the fermion mass term, because it does provide terms of the same order in $M$ as those in Eq.~(\ref{LUVJac}).
\end{itemize}

\subsection{EFTs with local and global symmetries}

The goal of the present paper is to consider scenarios combining both situations we have discussed so far, that is, with spontaneously broken gauge symmetries and anomalous global symmetries. Generically, our theory of interest corresponds to 
\begin{align}
\mathcal{L}_{\mathrm{UV}} &\supset \mathcal{L}_{\rm UV}^{\rm fermion} + \mathcal{L}_{\rm UV}^{\rm Jac}\, ,
\label{UVtot}
\end{align}
with
\begin{align}
\mathcal{L}_{\rm UV}^{\rm fermion} = \bar{\Psi} \bigg[ i\partial_{\mu}\gamma^{\mu
} - M & + \bigg( V_{\mu}-\dfrac{\partial_{\mu}\pi_V}{2v_V}\bigg)\gamma^{\mu} - \bigg( A_{\mu
}-\dfrac{\partial_{\mu}\pi_{A}}{2v_A}\bigg)\gamma^{\mu}\gamma^{5}
\nonumber \\
& - \bigg(0-\dfrac{\partial_{\mu}\pi_S}{2v_S}\bigg)\gamma^{\mu} - \bigg( 0 -\dfrac{\partial_{\mu}\pi_U}{2v_U}\bigg)\gamma^{\mu}\gamma^{5} \bigg]\Psi
\, .
\label{UVfermtot}
\end{align}
and for the Jacobian, using the singlet anomaly result for chiral fermions~\cite{Bilal:2008qx,Bertlmann:1996xk}, and noting that $\Psi_{L/R}$ couples to $V^\mu \pm A^\mu$ and $\partial_{\mu}(\pi_S \pm \pi_U)$,
\begin{align}
\mathcal{L}_{\rm UV}^{\rm Jac} & = 
\dfrac{1}{16\pi^{2}}\dfrac{\pi_{U}}{2v_{U}}\bigg[ \big(F_{_V,\mu\nu} + F_{_A,\mu\nu}\big)\big( \tilde{F}_{_V}^{\mu\nu} + \tilde{F}_{_A}^{\mu\nu} \big) + \big(F_{_V,\mu\nu} - F_{_A,\mu\nu}\big) \big( \tilde{F}_{_V}^{\mu\nu} - \tilde{F}_{_A}^{\mu\nu} \big) \bigg] 
\nonumber \\
& + \dfrac{1}{16\pi^{2}}\dfrac{\pi_{S}}{2v_{S}}\bigg[ \big(F_{_V,\mu\nu} + F_{_A,\mu\nu}\big)\big( \tilde{F}_{_V}^{\mu\nu} + \tilde{F}_{_A}^{\mu\nu} \big) - \big(F_{_V,\mu\nu}-F_{_A,\mu\nu}\big)\big( \tilde{F}_{_V}^{\mu\nu} - \tilde{F}_{_A}^{\mu\nu} \big)\bigg] \nonumber \\
& = \dfrac{1}{8\pi^{2}}\dfrac{\pi_{U}}{2v_U}\left(  F_{_{V},\mu\nu} \tilde{F}^{\mu\nu}_{_V} 
+ F_{_{A},\mu\nu}\tilde{F}_{_A}^{\mu\nu}\right)
+\dfrac{1}{4\pi^{2}}\dfrac{\pi_{S}}{2v_{S}}F_{_{A},\mu\nu}\tilde{F}_{_V}^{\mu\nu}
\, ,
\label{UVJactot}
\end{align}
where $\pi_{A}$ and $\pi_{V}$ have no contact interactions with field strength tensors since these gauge interactions are assumed anomaly-free. To insist on the fact that $\pi_{S}$ and $\pi_{U}$ are Goldstone bosons associated to global symmetries, we explicitly assign their respective would-be-gauge fields to $0$ in Eq.~\eqref{UVtot}. In this expression and throughout the rest of this section, we have set all the couplings to one to unclutter the derivation, but they can be straightforwardly reintroduced, as we will do in the following sections. 
This parametrisation of the fermion sector of the UV theory deserves several important comments:

\begin{itemize}
\item The UV theory necessarily involves several complex scalar fields, several species of fermions to cancel the gauge anomalies, along with some set of scalar and fermion couplings ensuring the existence of the gauge and global symmetries at the Lagrangian level. Further, as will be detailed in Sec.~4, the pseudoscalar components of these scalar fields in general mix, with some combinations eaten by the gauge fields, and some left over as true physical degrees of freedom. With the above parametrisation, we single out one of these fermions, and all the other UV features are encoded into the parameters $v_{S,U}$, $v_{A,V}$, which in general involves vacuum expectation values and some mixing angles, and in the fermion mass term $M$, which in general arises from several Yukawa couplings.

\item Adopting a non-linear representation for the scalar fields, with its associated loss of renormalisability, is inevitable if one wishes to leave the details of the whole scalar sector unspecified and start at the UV scale with an effective theory involving only the Goldstone bosons. Indeed, those have to be constrained to live on the specific coset space corresponding to the assumed symmetry breaking pattern. Note that for an abelian global symmetry, the dynamics of the Goldstone bosons is particularly simple, as there are no contact interactions among them, and all that remains is the shift symmetry.

\item One of the main goal of this work is to build an EFT by integrating out chiral fermions. As we have discussed, it is then convenient to reparametrise the fermion fields, so that the Goldstone boson couplings to fermions involve local partial derivative. This first ensures the gauge and shift symmetries are manifest, but it also makes the fermion mass term invariant under all symmetries. Though not compulsory, it then allows to construct the EFT by factoring the mass term out in a symmetry preserving way.

\item For the abelian toy model described here, the Goldstone bosons involved in vector currents, $\pi_S$ and $\pi_V$, actually play no role. Indeed, for the vector gauge interaction, the $\partial_{\mu}\pi_{V}\bar{\Psi}\gamma^{\mu}\Psi$ interaction can always be eliminated by a non-anomalous reparametrisation $\Psi\rightarrow\exp\big(i\frac{\pi_{V}}{2v_{V}}\big)\Psi$, which leaves the fermion mass term invariant. Whether it is spontaneously broken or not is thus irrelevant. For the scalar $\pi_S$ Goldstone boson associated to a global symmetry, the reparametrisation $\Psi\rightarrow\exp\big(i\frac{\pi_{S}}{2v_{S}}\big)\Psi$ not only removes the $\partial_{\mu}\pi_{S}\bar{\Psi}\gamma^{\mu}\Psi$ interaction, but being anomalous, it induces a Jacobian that precisely kills the $\pi_S$ terms in Eq.~(\ref{UVJactot}). The field $\pi_S$ thus disappears  entirely from the theory. These two facts are truly peculiar to the abelian gauge symmetry case, with the fermion in a one-dimensional representation. So, to set up the formalism to deal with more general theories, like the SM, we keep these fields explicitly in the UV parametrisation of the fermion couplings.\footnote{Further, integrating out the fermion starting from Eq.~(\ref{UVfermtot}), to verify that the $\pi_S$ derivative interaction indeed induces EFT operators that precisely cancel the Jacobian term in Eq.~(\ref{UVJactot}) provides a non-trivial check for our calculation, see Sec.~4.1.}
\end{itemize}

So, let us proceed and integrate out the fermion field involving local partial derivatives in its quadratic operator. Details of the calculation will be presented in the following section, but let us already discuss some interesting generic features. If one decides to use Feynman diagrams to integrate out fermions, one will have to deal with  divergent triangle amplitudes that one will have to carefully regularise. Even if this is a standard manipulation in QFT, the potential spread of the anomaly has to be considered with great care as discussed in Refs.~\cite{Quevillon:2019zrd, Quevillon:2020hmx}. 
In the functional approach, that we will follow all along this work, the fact that the axial vector or vector couplings are anomalous
manifests itself by the presence of ambiguities in the functional trace~\footnote{More precisely the ambiguity is localised in the Dirac matrices trace if one chooses to use dimensional regularisation, as we will do.}.

This means that starting from Eq.~\eqref{UVfermtot}, the fermion-less EFT expansion will start with six dimension-five operators involving the Goldstone bosons $\pi_U$ and $\pi_S$~\footnote{A priori the only non vanishing dimension five operators have to involve Dirac traces with only one $\gamma^{5}$ matrix or with three $\gamma^{5}$ matrices.},
\begin{align}
\mathcal{L}_{\rm UV}^{\rm fermion} \rightarrow \mathcal{L}_{\rm EFT}^{\rm 1loop}  & = \omega_{_{AVV}}\dfrac{\partial_{\mu}\pi_{U}}{2v_{U}}V_{\nu}\tilde{F}_{_V}^{\mu\nu}+\omega_{_{AAA}}\dfrac{\partial_{\mu}\pi_{U}}{2v_{U}}\left(  A_{\nu}-\dfrac{\partial_{\nu}\pi_{A}}{2v_{A}}\right)  \tilde{F}_{_A}^{\mu\nu}
\nonumber\\
& +\omega_{_{VVA}}\dfrac{\partial_{\mu}\pi_{S}}{2v_{S}}V_{\nu}\tilde{F}_{_A}^{\mu\nu}+\omega_{_{VAV}}\dfrac{\partial_{\mu}\pi_{S}}{2v_{S}}\left(
A_{\nu}-\dfrac{\partial_{\nu}\pi_{A}}{2v_{A}}\right)  \tilde{F}_{_V}^{\mu\nu}\ .
\label{initial4op}
\end{align}
The evaluation of the $\omega_i$ coefficients~\footnote{The $\omega_i$ coefficients carry the CP properties of their associated three Lorentz structures. As an example, $\partial_{\mu}\pi_{U}$ is CP-odd, $V_{\nu}$ and $\tilde{F}_{_V}^{\mu\nu}$ are CP-even so the associated coefficient reads $\omega_{AVV}$.} involves divergent integrals and after their regularisation, those parameters end up fully ambiguous. We thus need to find a strategy to fix them.

Actually, these ambiguities are the exact analog of those arising for the triangle diagrams, whose expressions are ambiguous since they depend on the routing of the momenta when working at the Feynman diagram level. In that case, the ambiguities are removed by imposing the appropriate Ward identities, that is, gauge invariance. So, we would like to do the same here, and impose the vector and axial gauge invariance. However, all the operators in Eq.~\eqref{initial4op} are already gauge invariant! Actually, the would-be-Goldstone bosons $\pi_{A}$ are not even needed to ensure the gauge invariance, and they never contribute to S-matrix elements. The reason is that their contributions, or the $\theta_{_{V,A}}$ terms arising when $V_{\mu}\rightarrow V_{\mu}+\partial_{\mu}\theta_{_V}$ or $A_{\mu}\rightarrow A_{\mu}+\partial_{\mu}\theta_{_A}$, drop out by integration by parts~\footnote{One should note that integration by parts can be performed without any hesitation since the fermion has been formally integrated out.} thanks to the antisymmetry of $\tilde{F}_{_{V,A}}^{\mu\nu}$ and the Bianchi identity.

In the initial Lagrangian of Eq.~\eqref{UVtot} we decided to treat both the would-be Goldstone bosons ($\pi_V$ and $\pi_A$) and the Goldstone bosons ($\pi_S$ and $\pi_U$) on equal footing by writing them with local derivative acting on them. Since this increases the degree of divergence of integrals one would then be tempted, in order to minimise the number of integrals to regularise, to preferentially consider the situation where the would-be Goldstone bosons enter the mass term (let us remind that this can be trivially done since the gauge symmetries are assumed not to be anomalous). Then, after Taylor expanding the mass term one obtains, 
\begin{align}
\mathcal{L}_{\rm UV}^{\rm fermion} = \bar{\Psi} \bigg[ i\partial_{\mu}\gamma^{\mu
} - M\bigg( 1 +\dfrac{\pi_{A}}{v_A}\,i\gamma^{5} \bigg)+ V_{\mu}\gamma^{\mu} - A_{\mu}\gamma^{\mu}\gamma^{5} + \dfrac{\partial_{\mu}\pi_S}{2v_S} \gamma^{\mu} + \dfrac{\partial_{\mu}\pi_{U}}{2v_U}\gamma^{\mu}\gamma^{5} \bigg]\Psi 
\, .
\label{UVtot2}
\end{align}
Since by construction the $U(1)_V$ and $U(1)_A$ symmetries are gauged, the would-be-Goldstone bosons, $\pi_V$ and $\pi_A$, are not involved in bosonic operators up to dimension five, starting from Eq.~\eqref{UVtot2}. This means that the fermion-less EFT expansion will start again with four dimension-five operators involving only the Goldstone bosons $\pi_S$ and $\pi_U$~, 
\begin{align}
\mathcal{L}_{\rm UV}^{\rm fermion} \rightarrow \mathcal{L}_{\rm EFT} & =  
\omega_{_{AVV}}\dfrac{\partial_{\mu}\pi_{U}}{2v_{U}}V_{\nu}\tilde{F}_{_V}^{\mu\nu}
+\omega_{_{AAA}}\dfrac{\partial_{\mu}\pi_{U}}{2v_{U}} A_{\nu} \tilde{F}_{_A}^{\mu\nu}  
\nonumber \\
& + \omega_{_{VVA}}\dfrac{\partial_{\mu}\pi_{S}}{2v_{S}}V_{\nu}\tilde{F}_{_A}^{\mu\nu} 
+ \omega_{_{VAV}}\dfrac{\partial_{\mu}\pi_{S}}{2v_{S}} A_{\nu}\tilde{F}_{_V}^{\mu\nu} \ . 
\end{align}

Since $\pi_A$ drops out of Eq.~\eqref{initial4op} under integration by part, we recover exactly the same effective interactions. Moving the would-be-Goldstone to the fermion mass term, that is, making it gauge-dependent, does not help to fix the $\omega_{i}$ coefficients because gauge invariance is still automatic for the leading dimension-five operators. The only way forward is to perturb the theory to break this automatic gauge invariance, so that non-trivial constraints on the $\omega_{i}$ coefficients can emerge. One possibility is to associate to the Goldstone bosons, $\pi_S$ and $\pi_U$, auxiliary gauge fields $S_{\mu}$ and $U_{\mu}$, respectively, as we will now discuss.

\subsection{Remove ambiguities with artificial gauging}

One way to fix $\omega_{_{AVV}}$, $\omega_{_{VVA}}$, $\omega_{_{VAV}}$ and $\omega_{_{AAA}}$ using the constraint of gauge invariance is to introduce fictitious~\footnote{At the end of the day, we will still want the global symmetry to stay global and to set to zero these fictitious vector fields.} vector and axial vector gauge fields associated to the $\pi_{S}$ and $\pi_{U}$ Goldstone bosons.  These fictitious gauge fields then enter in the effective operators of Eq.~\eqref{initial4op}, and prevent gauge invariance from being automatic under partial integration. They also prevent the contributions involving the would-be-Goldstone bosons of the true symmetries to vanish. This trick, introduced in Ref.~\cite{Bonnefoy:2020gyh}, is the key to derive non-trivial constraints and fix the ambiguous coefficients.

One may be a bit uneasy about this gauging of the global symmetries since these are precisely the symmetries that are anomalous. Actually, in the following, we will never need to use the fictitious gauge invariance in any form. All that matters is that these fictitious gauge fields act as background fields for $\partial_{\mu}\pi_{S}$ and $\partial_{\mu}\pi_{U}$, so as to upset the automatic (true) gauge invariances. This is sufficient to derive non-trivial constraints from the true, non-anomalous gauge symmetries.

Yet, as advocated in Ref.~\cite{Bonnefoy:2020gyh}, it can also be technically interesting to view these background fields as fictitious gauge fields, because then all the symmetries are treated on the same footing. As we will detail in section~\ref{UOLEA}, the calculation of the EFT becomes fully generic. The nice feature is that under this form, one can decide only at the very end which of the gauge symmetries is to be anomalous, hence fictitious, by imposing the exact invariance of the EFT under the \textit{other} gauge symmetries, those that are kept active. 

To illustrate all that, let us thus rewrite our initial Lagrangian as
\begin{align}
\mathcal{L}_{\rm UV, I}^{\rm fermion} = \bar{\Psi} \bigg[ i\partial_{\mu}\gamma^{\mu
} - M & +\bigg( V_{\mu}-\dfrac{\partial_{\mu}\pi_V}{2v_V}\bigg)\gamma^{\mu} - \bigg( A_{\mu
}-\dfrac{\partial_{\mu}\pi_{A}}{2v_A}\bigg)\gamma^{\mu}\gamma^{5}
\nonumber \\
& + \bigg(S_{\mu}-\dfrac{\partial_{\mu}\pi_S}{2v_S}\bigg)\gamma^{\mu} 
 + \bigg( U_{\mu} -\dfrac{\partial_{\mu}\pi_{U}}{2v_U}\bigg)\gamma^{\mu}\gamma^{5} \bigg]\Psi
\, .
\label{UVtotTrick1}
\end{align}
In this Lagrangian, the $\partial_{\mu}\pi_{V}$ piece is irrelevant, since it can be eliminated by an innocuous reparametrisation, but let us keep it anyway for now. Integrating out the fermion leads to the EFT :
\begin{align}
\mathcal{L}_{\rm EFT, I}^{\rm 1loop} & = 
 \omega_{_{VVA}}\left(  S_{\mu}-\dfrac{\partial_{\mu}\pi_{S}}{2v_{S}}\right)  \left(V_{\nu} -\dfrac{\partial_{\nu}\pi_{V}}{2v_{V}}\right) \tilde{F}_{_A}^{\mu\nu} + \omega_{_{AVV}}\left( U_{\mu} - \dfrac{\partial_{\mu}\pi_{U}}{2v_{U}}\right)  \left(V_{\nu} -\dfrac{\partial_{\nu}\pi_{V}}{2v_{V} }\right)\tilde{F}_{_V}^{\mu\nu} 
\nonumber\\
& + \omega_{_{VAV}}\left(  S_{\mu}-\dfrac{\partial_{\mu}\pi_{S}}{2v_{S}}\right) \left(A_{\nu} -\dfrac{\partial_{\nu}\pi_{A}}{2v_{A}}\right) \tilde{F}_{_V}^{\mu\nu} +\omega_{_{AAA}}\left(  U_{\mu}-\dfrac{\partial_{\mu}\pi_{U}}{2v_{U}}\right) \left(A_{\nu} -\dfrac{\partial_{\nu}\pi_{A}}{2v_{A} }\right)  \tilde{F}_{_A}^{\mu\nu} 
\, ,
\label{EfftotTrickk1}
\end{align}
with again the ambiguous coefficients $\omega_{_{AVV}}$, $\omega_{_{VVA}}$, $\omega_{_{VAV}}$ and $\omega_{_{AAA}}$ (the details of the calculation will be presented in the next section). All these interactions are still automatically gauge invariant thanks to the presence of the would-be-Goldstone bosons. Now, the key is to remember that the true gauge interactions are anomaly-free by assumption. This means the $\pi_A$ can be freely moved to the mass term by a reparametrisation of the fermion field, without Jacobian, and as said above, the $\partial_{\mu}\pi_{V}$ term can be discarded, again without Jacobian. Thus, the UV Lagrangian can equivalently be written as
\begin{align}
\mathcal{L}_{\rm UV, II}^{\rm fermion} = \bar{\Psi} \bigg[ i\partial_{\mu}\gamma^{\mu
} & - M\bigg( 1 + \dfrac{\pi_{A}}{v_A}\,i\gamma^{5} \bigg)+ V_{\mu}\gamma^{\mu} - A_{\mu}\gamma^{\mu}\gamma^{5}
\nonumber \\
& + \bigg(S_{\mu}-\dfrac{\partial_{\mu}\pi_S}{2v_S}\bigg)\gamma^{\mu} + \bigg( U_{\mu} -\dfrac{\partial_{\mu}\pi_{U}}{2v_U}\bigg)\gamma^{\mu}\gamma^{5} \bigg]\Psi
\, .
\label{UVtotTrick2}
\end{align}
This time, there is no ambiguity in calculating the Wilson coefficients of the operators involving the would-be-Goldstone bosons. The five-dimensional effective interactions become 
\begin{align}
\mathcal{L}_{\rm EFT, II}^{\rm 1loop} & \supset
\omega_{_{VVA}}\left(  S_{\mu}-\dfrac{\partial_{\mu}\pi_{S}}{2v_{S}}\right)  V_{\nu}\tilde{F}_{_A}^{\mu\nu} + \omega_{_{AVV}}\left(U_{\mu}-\dfrac{\partial_{\mu}\pi_{U}}{2v_{U}}\right)  V_{\nu}\tilde{F}_{_V}^{\mu\nu} \nonumber\\
& 
+\omega_{_{VAV}}\left( S_{\mu}-\dfrac{\partial_{\mu}\pi
_{S}}{2v_{S}}\right) A_{\nu} \tilde{F}_{_V}^{\mu\nu} + \eta_{_{ASV}} \dfrac{\pi_A}{v_A} F_{_S,\,\mu\nu} \tilde{F}_{_V}^{\mu\nu}   
\nonumber\\
&
+\omega_{_{AAA}}\left( U_{\mu}-\dfrac{\partial_{\mu}\pi_U}{2v_{U}}\right) A_{\nu} \tilde{F}_{_A}^{\mu\nu}  + \eta_{_{AUA}} \dfrac{\pi_A}{v_A} F_{_U,\,\mu\nu} \tilde{F}_{_A}^{\mu\nu} 
\, ,
\label{EfftotTrickk2}
\end{align}
where $\omega_{_{AVV}}$, $\omega_{_{VVA}}$, $\omega_{_{VAV}}$ and $\omega_{_{AAA}}$ are ambiguous, but not $\eta_{_{ASV}}$ and $\eta_{_{AUA}}$ since they arise from convergent integrals. Importantly, under this form, the true $U(1)_V$ and $U(1)_A$ gauge invariances are no longer automatic.

Now, we end up with two equivalent ways to fix the ambiguities. Either we enforce the matching of Eq.~(\ref{EfftotTrickk2}) with Eq.~(\ref{EfftotTrickk1}), or we impose gauge invariance on Eq.~(\ref{EfftotTrickk2}). In both cases, the constraints take the same form, but the latter is obviously more economical from a calculation point of view and will be adopted in the next sections.

For instance, for the vector gauge fields, since $\pi_{V}$ is absent from Eq.~(\ref{EfftotTrickk2}), matching with Eq.~(\ref{EfftotTrickk1}) requires $\omega_{_{VVA}}$ and $\omega_{_{AVV}}$ to vanish. Equivalently, invariance of Eq.~(\ref{EfftotTrickk2}) under  $V_{\mu}\rightarrow V_{\mu}+\partial_{\mu}\theta_{_V}$ immediately imposes $\omega_{_{VVA}}=\omega_{_{AVV}}=0$. This corresponds to the usual result that for vector gauge interactions, the derivative interactions of a Goldstone boson with the fermions contributes only at the subleading order in the mass expansion, otherwise known as the Sutherland-Veltman theorem. The local Jacobian terms in Eq.~\eqref{UVJactot} immediately catch the whole $\pi_{U}VV$ coupling.

For the axial gauge field, matching Eq.~(\ref{EfftotTrickk2}) with Eq.~(\ref{EfftotTrickk1}) obviously permits to fix the ambiguous $\omega_{_{VAV}}$ and $\omega_{_{AAA}}$ in terms of $\eta_{_{AVS}}$ and $\eta_{_{AUA}}$, which are fully calculable. Alternatively, performing a $U(1)_A$ gauge transformation $A_{\mu}\rightarrow A_{\mu}+\partial_{\mu}\theta_{_A}$ together with $\pi_{A}\rightarrow\pi_{A} + 2v_A\theta_{_A}$ in Eq.~(\ref{EfftotTrickk2}) generates the gauge variation, after integrating by part and using the Bianchi identity,  
\begin{align}
      \delta_A\big(\mathcal{L}_{\rm EFT, II}^{\rm 1loop}\big) &= \bigg( \dfrac{1}{2}\omega_{_{VAV}} + 2\eta_{_{ASV}} \bigg) \theta_{_A} F_{_S}^{\mu\nu}\tilde{F}_{_V,\,\mu\nu}
    + \bigg( \dfrac{1}{2} \omega_{_{AAA}} + 2\eta_{_{AUA}} \bigg) \theta_{_A} F_{_U}^{\mu\nu}\tilde{F}_{_A,\,\mu\nu}
    \, .
    \label{CoeffTrick1}
\end{align}
Hence, the requirement of gauge invariance asks for
\begin{align}
    \delta_A\big(\mathcal{L}_{\rm EFT, II}^{\rm 1loop}\big) &= 0 \Leftrightarrow \omega_{_{VAV}} = -4 \eta_{_{ASV}} ~ \text{and}~ \omega_{_{AAA}} = -4 \eta_{_{AUA}} 
    \, .
    \label{CoeffTrick2}
\end{align}
The effective axion-bosonic Lagrangian is obtained by adding $\mathcal{L}_{\rm UV}^{\rm Jac}$ and $\mathcal{L}_{\rm EFT,II}^{\rm 1loop}$ and finally setting the fictitious vector fields to zero, this gives the result,
\begin{align}
    \mathcal{L}_{\rm EFT} 
&= \dfrac{1}{16\pi^2}\dfrac{\pi_U}{v_U}F_{_{V},\mu\nu} \tilde{F}^{\mu\nu}_{_V} 
+ \bigg[\dfrac{1}{16\pi^2}-\eta_{_{AUA}}\bigg]\dfrac{\pi_U}{v_U}F_{_A,\mu\nu}\tilde{F}_{_A}^{\mu\nu} 
+ \bigg[\dfrac{1}{8\pi^2}-\eta_{_{ASV}}\bigg]\dfrac{\pi_S}{v_S}F_{_A,\mu\nu}\tilde{F}_{_V}^{\mu\nu}
\, .
\end{align}
Let us stress again that the $\eta_{_{AUA}}$ and $\eta_{_{ASV}}$ are fully calculable, unambiguous coefficients originating from convergent integrals. The determination of $\omega_{_{VAV}}$ and $\omega_{_{AAA}}$ from the requirement of gauge invariance is now transparent, and precisely matches that using Ward identities in a Feynman diagram context~\cite{Quevillon:2019zrd}. This is the general procedure we will adopt in the following to derive our bosonic EFTs. Of course, in the physical case, none of the interactions parametrised by $\eta_{_{ASV}}$ and $\eta_{_{AUA}}$ exist since they require the presence of the fictitious $U_{\mu}$ and $S_{\mu}$ gauge
fields as background values\footnote{Looking back, it is clear that gauge invariance under these fictitious symmetries is never imposed in any form. All that matters is to prevent the would-be-Goldstone bosons from being automatically absent from both Eq.~(\ref{EfftotTrickk1}) and Eq.~(\ref{EfftotTrickk2}), and true gauge invariance from being automatic in both EFT Lagrangians.}. Yet, this derivation sheds a new light on the violation of the Sutherland-Veltman theorem in the presence of spontaneously broken axial gauge interactions. Ultimately, it is due to the contribution of the associated would-be-Goldstone boson. The net effect is that the $\pi_{S}VA$ and $\pi_{U}AA$ couplings are not fully determined by the corresponding terms in the Jacobian, Eq.~\eqref{UVJactot}, since derivative interactions do contribute at leading order in the inverse mass expansion.

\section{Integrating out chiral fermions}\label{UOLEA}

In the previous section we discussed, qualitatively, peculiarities arising when building an EFT, while integrating out fermionic fields, from a UV theory with exact or spontaneous gauge symmetries and anomalous global symmetries. In this section we will, quantitatively, construct these EFTs involving gauge fields and their associated would-be-Goldstone bosons and simple Goldstone bosons associated to global symmetries. While the would-be Goldstone bosons can display derivative or pseudo-scalar couplings to fermions, since ultimately this depends on the fermion parametrisation (as we have discussed before), the Goldstone bosons will have to be taken firmly with local derivative couplings to fermions. Strictly speaking, from a path integral point of view, those details of the model are not mandatory to perform the main computation part, meaning forming the operator basis, evaluating the loop integrals after regularizing them. The symmetry aspects of the model will only matter at the very last stage when matching a UV theory onto its EFT.

In this section, we will briefly review the core techniques for calculating Wilson coefficients of EFT higher dimensional operators at the one-loop level by utilizing the functional approach. Since our interest is about the anomaly structure of specific QFTs, we will concentrate, in a general way, on the task of integrating out chiral fermions or fields which chiraly interact with gauge fields. We will also remind the reader how anomalies arise depending on how the one-loop effective action is regularised.

\subsection{Evaluation of the fermionic effective action}\label{effective_action}

We consider a generic UV theory containing a heavy Dirac fermion $\Psi$ of mass $M$ interacting bilinearly with a light field $\phi$, which is encapsulated inside the background function $\mathrm{X}[\phi]$ \footnote{For simplicity, we will consider $\Psi$ and $\phi$ as singlets but the following procedure is more general and it is still possible to treat them as multiplets.}.  The matter Lagrangian of this generic UV theory can be written as follows,
\begin{align}
\mathcal{L}_{\rm UV}^{\rm fermion}\big[ \Psi, \phi \big] 
&\supset \bar{\Psi} \bigg[ P_{\mu}\gamma^{\mu} - M + X[\phi]  \bigg]\Psi
= \bar{\Psi} \,  \mathcal{Q}_{\rm UV}[\phi] \,  \Psi \, ,
\label{uolea: UV-Lagrangian}
\end{align}
where $P_{\mu} = i\partial_{\mu}$ and introducing $\mathcal{Q}_{UV}[\phi]$ the fermionic quadratic operator. The background function $X[\phi]$ that we will consider throughout this paper is
\begin{align}
    X[\phi] & = V_{\mu}[\phi]\gamma^{\mu} - A_{\mu}[\phi]\gamma^{\mu}\gamma^5 - W_1[\phi]i\gamma^5
    \, ,
\end{align}
where we decompose $X[\phi]$ in terms of vector $V_{\mu}[\phi]$, axial-vector $A_{\mu}[\phi]$ and pseudo-scalar $W_1[\phi]$ structures\footnote{We note that $V_{\mu}[\phi]$, $A_{\mu}[\phi]$ and $W_1[\phi]$ do not contain any Dirac matrices or momentum variables $q_{\mu}$. The structures $V_{\mu}[\phi]$ and $A_{\mu}[\phi]$ can include gauge fields or local derivative of scalar fields.}, which are all the different types of interactions we will need to match our ``axion motivated'' UV theory to an EFT. In order to obtain the fermionic one-loop effective action, the light field $\phi$ is treated classically,  integrating out the fermion field $\Psi$ yields\footnote{The quantity $S_{\rm EFT}^{\rm 1loop}$ corresponds to the fermion 1PI action and it is formally divergent. We will discuss its gauge variation and its regularisation in the following.}
\begin{align}
e^{iS_{\rm EFT}^{\rm 1loop}[\phi]} &= \int \mathcal{D}\overline{\Psi}\mathcal{D}\Psi \, e^{iS_{\rm UV}[\Psi ,  \, \phi]} 
\nonumber\\
&\simeq e^{iS_{\rm UV}[ \Psi_c ,  \, \phi ]} \int \mathcal{D}\bar{\eta} \, \mathcal{D}\eta \, e^{i\int d^4 x \, \bar{\eta} \mathcal{Q}_{\rm UV}[\phi] \eta }
= e^{iS_{\rm UV}[ \Psi_c , \,  \phi ]} \,  \det \mathcal{Q}_{\rm UV}[\phi] \nonumber \\ 
& = e^{iS_{\rm UV}[\Psi_c ,  \,  \phi ]} e^{ \text{Tr} \ln \mathcal{Q}_{\rm UV}[\phi]  } \,  ,
\label{uolea: 1PI-action}
\end{align} 
where in the second line of Eq.~\eqref{uolea: 1PI-action} we have expanded the fermion fields around their classical background values, $\Psi = \Psi_c + \eta$ and performed the integration over the quantum fluctuations $\eta$.  Eventually, we have traded the functional determinant for the functional trace, $``\text{Tr}"$, running over the functional space and internal indices of the quadratic operator, $\mathcal{Q}_{\rm UV}[\phi]$. We therefore arrive at the one-loop effective action arising from integrating out a fermion:
\begin{align}
    S_{\rm EFT}^{\rm 1loop} &= -i \, \text{Tr}\ln 
    ( \slashed{P} - M + V_{\mu}[\phi]\gamma^{\mu} - A_{\mu}[\phi]\gamma^{\mu}\gamma^5 - W_1[\phi]i\gamma^5)
    \, .
\label{S1LEFT}    
\end{align}
Generally, in the functional space, one can write the quadratic operator as a function of position, $\hat{x}$, and momentum, $\hat{p}$, operators. Projecting onto position space, these operators become $\hat{x}=x$ and $\hat{p}_{\mu}=i\partial_{\mu}$. The standard initial step is to evaluate the trace over functional space by inserting the momentum eigenstate basis together with employing the canonical quantum mechanical trick of inserting the identity matrix, $\int d^4x \ket{x}\bra{x} = \mathds{1}$~\footnote{For the reader who would like to investigate in details the whole computation steps, we recommend Refs.\cite{Henning:2014wua,Henning:2016lyp,Fuentes-Martin:2016uol,Cohen:2019btp}.},
\begin{align} 
   S_{\rm EFT}^{\rm 1loop} &= -i\, \int \dfrac{d^4q}{(2\pi)^4} \bra{q} \text{tr} \ln \mathcal{Q}_{UV}(\hat{x},\hat{p}_{\mu}) \ket{q}
    \nonumber \\
    &= -i\, \int d^4x \int \dfrac{d^4q}{(2\pi)^4} \braket{q}{x}\bra{x} \text{tr} \ln \mathcal{Q}_{UV}(\hat{x},\hat{p}_{\mu}) \ket{q}
    \nonumber \\
    &= -i\, \int d^4x \int \dfrac{d^4q}{(2\pi)^4} \, e^{iq\cdot x} \, \text{tr} \ln \mathcal{Q}_{UV}(x, i\partial_{\mu}) e^{-iq\cdot x}
    \nonumber \\
    &= \int d^4x \int \dfrac{d^4q}{(2\pi)^4} (-i) \, \text{tr} \ln \mathcal{Q}_{UV}(x, i\partial_{\mu} - q_{\mu}) \, ,
    \label{Tr-Q}
\end{align}
where ``tr" now denotes the trace over spinor and internal symmetry indices only. Here the $\bra{x}$ denotes the eigenstate of local operator in position space, e.g. $\bra{x}\mathcal{Q}_{UV}(\hat{x},\hat{p})= \mathcal{Q}_{UV}(x,i\partial_{\mu})\bra{x}$, and the convention for inner product is $\braket{x}{q}=e^{-iq\cdot x}$. An ``open" derivative from the kinematic operator will get shifted due to $e^{iq\cdot x} i\partial_{\mu} e^{-iq\cdot x} = i\partial_{\mu} +q_{\mu}$. We perform also a conventional change of integration variable $q \rightarrow -q$.
As we will study later, we emphasise that in the case where one has to deal with a local derivative of a bosonic field, e.g. $\big[\partial_{\mu}\pi(x) \big]$, this term will not be shifted under the sandwich of $e^{iq\cdot x}\big[ \partial_{\mu}\pi(x) \big]e^{-iq\cdot x}$ since the partial derivative of this coupling is ``closed". Therefore, on the computational side, depending on the vector or axial-vector nature of the local derivative couplings, one can absorb these terms into the vector ($V_{\mu}[\phi]$) and axial-vector ($A_{\mu}[\phi]$) structures of the UV quadratic operator~\footnote{This underlines the practical usefulness of our initial choice of parametrisation made in Eq.~\eqref{S1LEFT}}.

Ultimately, the expansion of the logarithm in terms of a series of local operators suppressed by the fermion mass scale can be performed by a variety of techniques,
\begin{align}
\mathcal{L}_{\rm EFT}^{\rm 1loop} &= -i \, \text{Tr}\ln 
    ( \slashed{P} - \slashed{q} - M - X[\phi]) \nonumber \\
&= i \, \tr \sum_{n=1}^{\infty} \dfrac{1}{n} \int \dfrac{d^4q}{(2\pi)^4} \left[ \,  \dfrac{-1}{\slashed{q}+M} \bigg( -\slashed{P} -V_{\mu}[\phi]\gamma^{\mu} + A_{\mu}[\phi] \gamma^{\mu}\gamma^5 + W_1[\phi]i\gamma^5 \bigg)  \right]^n \, .
\label{Tr-Q-expansion}
\end{align}
The remarkable point at this stage is that the q-momentum integration can be factorised out from the generic operator structures. Indeed, regardless of the method used to evaluate the logarithm expansion, it can be done once-and-for-all, and the result is the same and universal in the sense that the final expression is independent of the details of the UV Lagrangian, which remain encapsulated in the X matrix of light fields, covariant derivative $P_\mu$, and mass matrix $M$. This leads to the so-called concept of the Universal One-Loop Effective Action (UOLEA) (see Refs~\cite{Drozd:2015rsp,Zhang:2016pja,Ellis:2016enq,Ellis:2017jns,Ellis:2020ivx}). 

Note that in our calculations we will deal with multiple vector, axial vector and pseudo-scalar interactions so we will consider in all generalities
\begin{align}
    V_{\mu}[\phi] \equiv g_{_V}^i V_{\mu}^i[\phi^i]\, , ~
    A_{\mu}[\phi] \equiv g_{_A}^i A_{\mu}^i[\phi^i]\, , ~
    W_1[\phi] \equiv g_{_{W_1}}^i W_1^i[\phi^i] \, ,
\end{align}
with an implicit summation over the i index.

\subsection{Ambiguities and regularisation of the functional trace}\label{Ambiguous_traces}

The evaluation of the one-loop effective Lagrangian Eq.~\eqref{Tr-Q-expansion} usually encounters divergent integrals and we use dimensional regularisation~\cite{tHooft:1972tcz} to evaluate them along with the $\overline{MS}$ scheme for renormalisation. The traces over Dirac matrices have to be performed in $d=4-\epsilon$ dimensions, and the $\epsilon$-terms resulting from the contractions with the metric tensor (satisfying then $g^{\mu\nu}g_{\mu\nu}=d$) must be kept in the computations. These $\epsilon$-terms will then multiply with the $(1/\epsilon)$ pole of the divergent integrals and yield finite contributions. We emphasise that depending on the regularisation scheme for $\gamma^5$ in $d$-dimensions, different results for $\epsilon$-terms in Dirac traces will emerge (see for examples Refs.~\cite{Chanowitz:1979zu,Novotny:1994yx,Belusca-Maito:2020ala}). We will come back shortly to describe in details the prescription we used to evaluate ill-defined Dirac traces involving $\gamma^5$ matrices, in dimensional regularisation.
 
We now turn back on the ambiguities arising in some of our integrals in the 4-dimensional space. Usually, when computing one-loop divergent triangle Feynman diagrams (corresponding to the Adler–Bell–Jackiw anomaly \cite{Adler:1969gk,Bell:1969ts}), it is well-known that, in $d=4$ dimensions, an ambiguity of the loop integral arises. It corresponds to an arbitrariness in the chosen integration variables (see Ref.~\cite{Weinberg:1996kr}), and actually there can be surface terms that do depend on the chosen momentum routing. Those surface terms then contribute to the divergence of vector-currents and axial-vector-currents, and  all the naive Ward identities cannot be satisfied simultaneously. At least some of them will be anomalous. The important point is that the arbitrariness of integral variable can be parametrised in terms of free parameters (see the standard Refs.\cite{Weinberg:1996kr,Bilal:2008qx} and the more recent Refs.\cite{Quevillon:2019zrd,Bonnefoy:2020gyh}). By tuning the value of those free parameters, one can decide which symmetry is broken at the quantum level, and which are kept active. Evidently, to obtain the correct physical results, all the gauge symmetries must be preserved.

When switching to the d-dimensional space, the ambiguity on the loop integrals does not arise anymore from dependencies on the chosen momentum routing, but it is now inherent from the Dirac algebra sector. Indeed, not all the usual properties of the Dirac matrices can be maintained once in $d>4$ dimensions, essentially because $\gamma^5$ and the anti-symmetric tensor $\epsilon^{\mu\nu\rho\sigma}$ are intrinsically four-dimensional objects. Whatever the chosen definition, there is no way to consistently preserve both the anticommutativity properties of $\gamma^5$ matrices, i.e. $\{\gamma^{\mu},\gamma^5\}=0$, and the trace cyclicity property in $d>4$ dimensions. In the original work of 't Hooft and Veltman~\cite{tHooft:1972tcz}, they noted that the momentum routing ambiguity is replaced by an ambiguity in the location of $\gamma^5$ in the Dirac traces. Using their prescriptions for the Dirac algebra in $d>4$ dimensions (see Refs.\cite{tHooft:1972tcz,Breitenlohner:1977hr}), it is then possible to introduce free parameters keeping track of all the possible $\gamma^5$ locations in a given Dirac matrices string~\cite{Elias:1982ea}. As before, one can then tune these parameters to choose which symmetry is broken anomalously, and which one have to be preserved. This is the strategy we will employ to calculate the ambiguous Dirac traces in Eq.~\eqref{Tr-Q-expansion}.

\subsection{Evaluation of the anomaly related operators}{\label{GCS+Goldstone}}

We now concentrate on the derivation of the operators which ultimately involve a mixture of three gauge fields and Goldstone bosons with a derivative acting on them. With our parametrisation, they arise from combinations of the generic vector $V_{\mu}[\phi]$ and axial-vector $A_{\mu}[\phi]$ fields. Due to the presence of $\gamma_5$ Dirac matrices in their Wilson coefficients, they are truly ambiguous in dimensional regularisation.
Then we will proceed with the evaluation of operators involving one Goldstone boson (without any derivative acting on it), namely the $W_1[\phi]$ field in our generic parametrisation. 
These operators have been evaluated using the usual Feynman diagrams technique (see Refs.\cite{Quevillon:2019zrd,Bonnefoy:2020gyh,Anastasopoulos:2006cz}). Since those computations are subtle and lead to confusions, this is legitimate to wonder how one would perform them from a different point of view, such as within the path integral formalism. Which is what we present now.

\subsubsection{Evaluation of the ambiguous terms}

We start with the exercise of computing the divergent terms that naturally arise when evaluating Eq.~\eqref{Tr-Q-expansion}. The generic form of these operators is 
\begin{align}
    G_{\mu}^i G_{\nu}^j \tilde{F}_{\mu\nu}^k = G_{\mu}^i G_{\nu}^j \bigg( \dfrac{1}{2}\epsilon^{\mu\nu\rho\sigma}\partial_{_{[\rho}}G_{_{\sigma]}}^k \bigg)
    \, ,
    \label{GCS form}
\end{align}
where we use the notation $G_{\mu}^i$ to denote a generic gauge field and to avoid confusions with the vector and axial-vector structures in Eq.~\eqref{Tr-Q-expansion}. We also introduce the upper indices $i$,$j$,$k$ to keep the computation as general as possible and offer us the possibility to apply this computations to multiple gauge field configurations later on. Since starting with Eq.~\eqref{Tr-Q-expansion}, we chose to deal with vector and axial-vector structures, in order to reconstruct the ambiguous operators in the EFT, we need
\begin{itemize}
    \item One insertion of $P_{\mu}$ to account for the partial derivative and then allow to form a field strength tensor.
    \item Several combinations of vector and axial-vector structures. It is clear that to generate the anti-symmetric tensor $\epsilon^{\mu\nu\rho\sigma}$ the product of Dirac matrices must involve an odd number of $\gamma^5$ matrix. It exists only two possibilities, either an ``$AVV$'' contribution with one $\gamma^5$ or an ``$AAA$'' contribution with three $\gamma^5$.
\end{itemize}
While evaluating the one-loop effective Lagrangian of Eq.~\eqref{Tr-Q-expansion}, several contributions to the ambiguous effective interaction would arise from the $n=4$ polynomial terms 
\begin{align}
    \mathcal{L}_{\rm EFT}^{\rm 1loop} &\supset i \, \tr \, \dfrac{1}{4} \int \dfrac{d^dq}{(2\pi)^d} \left[ \,  \dfrac{-1}{\slashed{q}+M} \bigg( -P_{\mu}\gamma^{\mu} -V_{\mu}[\phi]\gamma^{\mu} +  A_{\mu}[\phi] \gamma^{\mu}\gamma^5 + W_1[\phi]\gamma^5 \bigg)  \right]^4
    \nonumber \\
    &\supset \sum_N f_{_N}^{^{AVV}}\mathbb{O}^{(PAVV)} + f_{_N}^{^{AAA}}\mathbb{O}^{(PAAA)}
    \, ,
    \label{GCS: EFT expansion}
\end{align}
where $\mathbb{O}^{(PAVV)}$ denotes the class of operator containing one $\gamma^5$ matrix and $\mathbb{O}^{(PAAA)}$ the one containing three $\gamma^5$ matrices.

\paragraph*{Evaluation of the $\mathbb{O}^{(PAVV)}$ structures.} There are three different types of combination that contribute to the $\mathbb{O}^{(PAVV)}$ structure, namely $\mathcal{O}(VVPA),\,\mathcal{O}(VAPV),\,\mathcal{O}(AVPV)$. Each type of combination contains four universal structures, and it is related together via trace cyclicity. At this stage, due to the ambiguity of $\gamma^5$-positions, one should not use trace cyclicity to minimise the number of universal structures that need to be evaluated. To present in detail the evaluation procedure of the Dirac trace and its regularisation, let us focus on one explicit example out of 12 universal structures included in Eq.~\eqref{GCS: EFT expansion}
\begin{align}
\mathcal{O}(VVPA) &\supset \dfrac{1}{4}  \int \dfrac{d^dq}{(2\pi)^d} \, \tr \bigg[  \dfrac{-1}{\slashed{q}+M} V_{\mu}\gamma^{\mu} \dfrac{-1}{\slashed{q}+M} V_{\nu}\gamma^{\nu} \dfrac{-1}{\slashed{q}+M} P_{\rho}\gamma^{\rho} \dfrac{-1}{\slashed{q}+M} A_{\sigma} \gamma^{\sigma}\gamma^5  \bigg]
\nonumber \\
&= \dfrac{i}{4} \bigg[ -4M^4 \Integral_i^4 + 16M^2 \Integral[q^2]_i^4 \bigg] \tr \bigg( \epsilon^{\mu\nu\rho\sigma} V_{\mu} V_{\nu} P_{\rho} A_{\sigma}  \bigg)
\nonumber \\
& + \dfrac{1}{4} \, \Integral[q^4]_i^4  \bigg[ g^{ab}g^{cd} + g^{ac}g^{bd} + g^{ad}g^{bc} \bigg]  \tr \bigg( \gamma_a \gamma^{\mu} \gamma_b \gamma^{\nu} \gamma_c \gamma^{\rho} \gamma_d \gamma^{\sigma} \gamma^5  \bigg) \bigg( V_{\mu} V_{\nu} P_{\rho} A_{\sigma}  \bigg)
\, ,
\label{example: avv-CDE-technique}
\end{align} 
where the fermion propagators are decomposed into $\dfrac{-1}{\slashed{q}+M} = \dfrac{M}{q^2-M^2} + \dfrac{-\slashed{q}}{q^2-M^2}$. For the tensorial integrals, we use
\begin{align}
         \int \dfrac{d^dq}{(2\pi)^d} \dfrac{q^{\mu_1}\cdots q^{\mu_{2n_c}}}{(q^2-m_i^2)^{n_i}(q^2-m_j^2)^{n_j}\cdots }
&= g^{\mu_1 \cdots \mu_{2n_c}} \mathcal{I}[q^{2n_c}]^{n_i n_j \cdots}_{i j \cdots } \, ,
        \label{formula: master-Integrals}
\end{align}
where $g^{\mu_1 \cdots \mu_{2n_c}}$ is the completely symmetric tensor, e.g. $g^{\mu\nu\rho\sigma} = g^{\mu\nu}g^{\rho\sigma}+g^{\mu\rho}g^{\nu\sigma}+g^{\mu\sigma}g^{\nu\rho}$, and we denote the master integrals as $\mathcal{I}[q^{2n_c}]^{n_i n_j \cdots}_{i j \cdots }$. The explicit expression and the value of some useful master integrals are derived in the Appendix \ref{Appendix:master_integrals}. In the second line of Eq.~\eqref{example: avv-CDE-technique}, all the loop integrals are finite, one can then evaluate the various Dirac traces in the usual naive scheme. The last line of Eq.~\eqref{example: avv-CDE-technique} contains divergent integrals, $\Integral[q^4]_i^4$, which have to be regularised. Let us show how to evaluate such an ambiguous quantity as $\tr \big(\gamma_a \gamma^{\mu}\gamma_b\gamma^{\nu}\gamma_c\gamma^{\rho}\gamma_d\gamma^{\sigma}\gamma^5 \big)$ of Eq.~\eqref{example: avv-CDE-technique}. We follow the procedure described earlier in the section \ref{Ambiguous_traces}. Before evaluating the Dirac trace in $d$-dimension, we first write down all possible structures that are equivalent to the original Dirac string by naively anti-commuting $\gamma^5$, 
\begin{align}
    \tr \big(\gamma_a \gamma^{\mu}\gamma_b\gamma^{\nu}\gamma_c\gamma^{\rho}\gamma_d\gamma^{\sigma} \gamma^5 \big) 
    &\rightarrow 
    \bar{a}_1 \tr \big(\gamma_a \gamma^{\mu}\Red{\gamma^5} \gamma_b\gamma^{\nu}\gamma_c\gamma^{\rho}\gamma_d\gamma^{\sigma}  \big)
    + \bar{a}_2 \tr \big(\gamma_a \gamma^{\mu}\gamma_b\gamma^{\nu}\Red{\gamma^5}\gamma_c\gamma^{\rho}\gamma_d\gamma^{\sigma} \big) 
    \nonumber \\
    &\quad + \bar{a}_3 \tr \big(\gamma_a \gamma^{\mu}\gamma_b\gamma^{\nu}\gamma_c\gamma^{\rho}\Red{\gamma^5}\gamma_d\gamma^{\sigma} \big) 
    + \bar{a}_4 \tr \big(\gamma_a \gamma^{\mu}\gamma_b\gamma^{\nu}\gamma_c\gamma^{\rho}\gamma_d\gamma^{\sigma}\Red{\gamma^5} \big) 
    \, ,
    \label{example: avv ambiguous traces}
\end{align}
where we introduce the four free parameters, $\bar{a}_i$, to keep track of the position of the $\gamma_5$ matrix in Eq.~\eqref{example: avv ambiguous traces}. Let us briefly comment on the fact that
\begin{itemize}
    \item In $d=4$ dimensions, all Dirac structures on the R.H.S of Eq.~\eqref{example: avv ambiguous traces} are equivalent.
    \item In $d=4-\epsilon$ dimensions, by using the Breitenlohner-Maison-’t Hooft-Veltman (BMHV) scheme (see Refs.\cite{tHooft:1972tcz,Breitenlohner:1977hr}), the $\gamma^5$ matrix does not anti-commute anymore with Dirac $\gamma^{\mu}$ matrices. Therefore, each Dirac trace will give a different result due to the different position of $\gamma^5$ matrix. The free parameters $\bar{a}_i$ is a device to keep track of the $\gamma^5$-positions.
    \item Enforcing a consistent result in $d=4$ and $d=4-\epsilon$ dimensions requires that $\sum_{i=1}^4\bar{a}_i=1$.
\end{itemize}
After plugging Eq.~\eqref{example: avv ambiguous traces} into  Eq.~\eqref{example: avv-CDE-technique}, one obtains
\begin{align}
\mathcal{O}(VVPA)
&\supset \dfrac{1}{4} \bigg[ 4M^4\Integral_i^4 - 16M^2 \Integral[q^2]_i^4 - 24\,\epsilon\,(-\bar{a}_1 + \bar{a}_2 - \bar{a}_3 + \bar{a}_4)\, \Integral[q^4]_i^4 \,  \bigg] \tr \bigg( \epsilon^{\mu\nu\rho\sigma} V_{\mu} V_{\nu} P_{\rho} A_{\sigma}  \bigg) 
\nonumber \\
&= \dfrac{1}{32\pi^2} \bigg[ -1 -\bar{a}_1 + \bar{a}_2 - \bar{a}_3 + \bar{a}_4 \, \bigg] \tr \bigg[ V_{\mu}^iV_{\nu}^j \tilde{F}_{\mu\nu}^{A^k} \bigg]
\, ,
\label{example: avv-Tr-D-dim}
\end{align}
where in the last line of Eq.~\eqref{example: avv-Tr-D-dim}, we replace the vector and axial-vector structures by $V_{\mu} \equiv g_{_V}^iV_{\mu}^i\,,~ A_{\mu} \equiv g_{_A}^iA_{\mu}^i \,$, we also omit the gauge couplings to simplify the expression of \eqref{example: avv-Tr-D-dim} and highlight the final value of loop integrals and Dirac traces. We remind the reader that $g_{_V}^i$ and $g_{_A}^i$ will only appear when it is necessary. Also, keep in mind that in Eq.~\eqref{example: avv-Tr-D-dim} the $\epsilon$-terms will hit the pole $\frac{1}{\epsilon}$ of the divergence integral, $\Integral[q^4]_i^4$, and generate finite contributions. We then apply the same method for the other contributions in $\mathbb{O}^{PAVV}$. One should note that since in Eq.~\eqref{Tr-Q-expansion}, $P_{\mu}=i\partial_{\mu}$, is the ``open" derivative one can therefore omit the operator structures which start with a $P_{\mu}$ since they lead to inert boundary terms. We underline one more time that at this stage, one cannot use the cyclicity property of the trace to reduce the number of terms that need to compute. Adding all the different contributions together gives
\begin{align}
\mathcal{L}_{\rm EFT}^{\rm 1loop} 
&\supset i\big( \, 24\epsilon\,\bar{a}_{V^iV^jA^k} \,  \Integral [q^4]_i^4 \,\big) \tr \bigg[  V_{\mu}^i V_{\nu}^j \tilde{F}_{\mu\nu}^{A^k} \bigg]
\nonumber \\
&+ i\big( -4M^4\Integral_i^4 + 16M^2\Integral[q^2]_i^4 + 24\epsilon\,\bar{a}_{V^jA^kV^i} \,\Integral[q^4]_i^4 \,\big) \tr \bigg[ V_{\mu}^j A^k_{\nu} \tilde{F}_{\mu\nu}^{V^i} \bigg] 
\nonumber\\
&+ i\big( 4M^4\Integral_i^4 - 16M^2\Integral[q^2]_i^4 + i\, 24\epsilon\,\bar{a}_{A^kV^iV^j} \,\Integral[q^4]_i^4 \,\big) \tr \bigg[ A^k_{\mu} V_{\nu}^i \tilde{F}_{\mu\nu}^{V^j} \bigg] 
\, .
\label{AVV: CDE-result}
\end{align}
Since the $\bar{a}_{i}$ coefficients are basically free, there are no reasons to give any physical meaning to the different contributions. For each operator structure, we redefine the total values of $\bar{a}_i$ by the new free parameters, e.g. $\bar{a}_{V^iV^jA^k}\,,~ \bar{a}_{V^jA^kV^i}\,,~ \bar{a}_{A^kV^iV^j}$. Readout the value of loop integrals, the above equation reduces to
\begin{align}
\mathcal{L}_{\rm EFT}^{\rm 1loop} 
&\supset \dfrac{1}{8\pi^2} \bar{a}_{V^iV^jA^k} \tr \bigg[  V_{\mu}^i V_{\nu}^j \tilde{F}_{\mu\nu}^{A^k} \bigg]
+ \dfrac{1}{8\pi^2} \bar{a}_{V^jA^kV^i} \tr \bigg[ V_{\mu}^j A^k_{\nu} \tilde{F}_{\mu\nu}^{V^i} \bigg] 
+ \dfrac{1}{8\pi^2} \bar{a}_{A^kV^iV^j} \tr \bigg[ A^k_{\mu} V_{\nu}^i \tilde{F}_{\mu\nu}^{V^j} \bigg] 
\, .
\label{AVV: CDE-result2}
\end{align}
The three operators of Eq.~\eqref{AVV: CDE-result2} are not independent and by using integration by parts one should always end up with two independent operators and then two free parameters.
As we will see later, in practice one decides to remove such or such operator by use of integration by parts based on the symmetries that are preserved or not since all operators are not invariant under the same vector or axial symmetries. As an example, if one supposes that, within our notation, the $V^i$ current might be anomalous, one may integrate by parts the first operator of Eq.~\eqref{AVV: CDE-result}, $\tr \big(V_{\mu}^i V_{\nu}^j \tilde{F}_{\mu\nu}^{A^k} \big)$, and after discarding the total derivative operator, and redefining the free parameters, one obtains
\begin{align}
\mathcal{L}_{\rm EFT}^{\rm 1loop} 
&\supset \dfrac{1}{8\pi^2} \bar{a}_{V^jA^kV^i} \tr \bigg[ V_{\mu}^j A^k_{\nu} \tilde{F}_{\mu\nu}^{V^i} \bigg] 
+ \dfrac{1}{8\pi^2}\bar{a}_{A^kV^iV^j} \tr \bigg[ A^k_{\mu} V_{\nu}^i \tilde{F}_{\mu\nu}^{V^j} \bigg] 
\, .
\label{AVV: CDE-result3}
\end{align}
At this point, one should comment on the fact that if one would have used the BMHV scheme without performing the decomposition of Eq.~\eqref{example: avv ambiguous traces}, one would have found each Wilson coefficients of the operators in Eq.~\eqref{AVV: CDE-result} to vanish. This is ultimately due to the fact that, by default, vector currents cannot be anomalous while only following the BMHV procedure. 
Even if one would have expected to be able to write effective operators as displayed in Eq.~\eqref{AVV: CDE-result2} from the first principle, we have rigorously shown how to obtain it in dimensional regularisation, i.e the ``AVV'' interaction can be described by two independent operators for which it exists two Wilson coefficients which are ambiguous i.e free. 

\paragraph*{Evaluation of the $\mathbb{O}^{(PAAA)}$ structures.} We now turn to the second class of operator,  $\mathbb{O}^{(PAAA)}$, that contains three $\gamma^5$ matrices. Similarly to the previous case with $\mathbb{O}^{(PAVV)}$, we start here by giving an explicit example for an operator that belongs to this class, 
\begin{align}
\mathbb{O}^{AAPA} &\supset \dfrac{1}{4}  \int \dfrac{d^dq}{(2\pi)^d} \, \tr \left[  \dfrac{-1}{\slashed{q}+M} A_{\mu} \gamma^{\mu} \gamma^5 \dfrac{-1}{\slashed{q}+M} A_{\nu} \gamma^{\nu}\gamma^5 \dfrac{-1}{\slashed{q}+M} P_{\rho}\gamma^{\rho} \dfrac{-1}{\slashed{q}+M} A_{\sigma} \gamma^{\sigma}\gamma^5  \right]^4
\nonumber \\
&= i\, \dfrac{1}{4} \bigg[ 4M^4 \Integral_i^4 + 16M^2 \Integral[q^2]_i^4 \bigg] \tr \bigg( \epsilon^{\mu\nu\rho\sigma} A_{\mu} A_{\nu} P_{\rho} A_{\sigma}  \bigg)
\nonumber \\
& + \dfrac{1}{4} \, \Integral[q^4]_i^4 \bigg[ g^{ab}g^{cd} + g^{ac}g^{bd} + g^{ad}g^{bc} \bigg] \tr  \bigg( \gamma_a \gamma^{\mu} \gamma^5 \gamma_b \gamma^{\nu} \gamma^5 \gamma_c \gamma^{\rho} \gamma_d \gamma^{\sigma} \gamma^5  \bigg) \bigg( A_{\mu} A_{\nu} P_{\rho} A_{\sigma} \bigg)
\, ,
\label{example: aaa Tr-ambiguous}
\end{align}
we then parameterise the ambiguous Dirac trace, $\tr \big(\gamma_a \gamma^{\mu}\gamma^5\gamma_b\gamma^{\nu}\gamma^5\gamma_c\gamma^{\rho}\gamma_d\gamma^{\sigma} \gamma^5 \big)$, by using
\begin{align}
    \tr \big(\gamma_a \gamma^{\mu}\gamma^5\gamma_b\gamma^{\nu}\gamma^5\gamma_c\gamma^{\rho}\gamma_d\gamma^{\sigma} \gamma^5 \big) 
    &\rightarrow 
    \bar{b}_1 \tr \big(\gamma_a \gamma^{\mu}\Red{\gamma^5} \gamma_b\gamma^{\nu}\gamma_c\gamma^{\rho}\gamma_d\gamma^{\sigma}  \big)
    + \bar{b}_2 \tr \big(\gamma_a \gamma^{\mu}\gamma_b\gamma^{\nu}\Red{\gamma^5}\gamma_c\gamma^{\rho}\gamma_d\gamma^{\sigma} \big) 
    \nonumber \\
    &\quad + \bar{b}_3 \tr \big(\gamma_a \gamma^{\mu}\gamma_b\gamma^{\nu}\gamma_c\gamma^{\rho}\Red{\gamma^5}\gamma_d\gamma^{\sigma} \big) 
    + \bar{b}_4 \tr \big(\gamma_a \gamma^{\mu}\gamma_b\gamma^{\nu}\gamma_c\gamma^{\rho}\gamma_d\gamma^{\sigma} \Red{\gamma^5} \big) 
    \, .
    \label{example: aaa ambiguous traces}
\end{align}
Afterwards, evaluating in $d=4-\epsilon$ dimensions with BMHV's scheme, we obtain
\begin{align} 
\mathbb{O}^{AAPA} &\supset \dfrac{i}{4} \bigg[ 4M^4\Integral_i^4 + 16M^2 \Integral[q^2]_i^4 + 24\,\epsilon\, \big(-\bar{b}_1 + \bar{b}_2 -\bar{b}_3 + \bar{b}_4 \big)\, \Integral[q^4]_i^4 \, \bigg] \tr \bigg( \epsilon^{\mu\nu\rho\sigma} A_{\mu} A_{\nu} P_{\rho} A_{\sigma}  \bigg) \nonumber \\
&= \dfrac{1}{32\pi^2}\bigg[ \dfrac{1}{3} +\bar{b}_1 - \bar{b}_2 +\bar{b}_3 - \bar{b}_4 \bigg] \tr \bigg[ A_{\mu}^iA_{\nu}^j\tilde{F}_{\mu\nu}^{A^k} \bigg]
\, ,
\label{example: aaa-Tr-D-dim}
\end{align}
where in the last step of the computation we evaluate the value of loop integrals, express $A_{\mu}\equiv A_{\mu}^i$. We also note that $g_{_A}^i$ will appear when it is necessary.
The computation for the other operators belonging to $\mathbb{O}^{(PAAA)}$ are similar and the full result reads
\begin{align}
\mathcal{L}_{\rm EFT}^{\rm 1loop} 
&\supset \big( i\, 24\epsilon \bar{b}_{A^iA^jA^k} \,  \Integral[q^4]_i^4 \, \big) \tr \bigg[ A^i_{\mu}A^j_{\nu} \tilde{F}_{\mu\nu}^{A^k}  \bigg]
+ \big( i \,  24\epsilon \bar{b}_{A^jA^kA^i} \,  \Integral[q^4]_i^4 \,\big) \tr \bigg[ A^j_{\mu}A^{k}_{\nu} \tilde{F}_{\mu\nu}^{A^i}  \bigg]
\nonumber\\
&\quad + \big( i \,  24\epsilon \bar{b}_{A^kA^iA^j} \,  \Integral[q^4]_i^4 \,\big) \tr \bigg[ A^{k}_{\mu}A^i_{\nu} \tilde{F}_{\mu\nu}^{A^j}  \bigg] \, .
\end{align}
which basically resumes to
\begin{align}
\mathcal{L}_{\rm EFT}^{\rm 1loop} 
&\supset \dfrac{1}{8\pi^2} \bar{b}_{A^iA^jA^k} \tr \bigg[ A^i_{\mu}A^j_{\nu} \tilde{F}_{\mu\nu}^{A^k}  \bigg] +\dfrac{1}{8\pi^2} \bar{b}_{A^jA^kA^i} \tr \bigg[ A^j_{\mu}A^k_{\nu} \tilde{F}_{\mu\nu}^{A^i}  \bigg] 
+ \dfrac{1}{8\pi^2} \bar{b}_{A^kA^iA^j} \tr \bigg[ A^k_{\mu}A^{i}_{\nu} \tilde{F}_{\mu\nu}^{A^j}  \bigg] \, .
\label{AAA: CDE-result3}
\end{align}

These three operators in Eq.~\eqref{AAA: CDE-result3} are not independent and one is free to remove one by the use of integration by parts. Consequently, in dimensional regularisation, the ``AAA'' interaction can be described by two independent operators attached to two free Wilson coefficients reflecting ambiguities in the evaluations of such interactions.

\subsubsection{Evaluation of the pseudo-scalar unambiguous terms}

We are now looking for to evaluate operators involving a pseudo-scalar $\phi$ (without local partial derivative acting on it) and two field strength tensors. The generic operator form is given by
\begin{align}
    \phi \, F_{\mu\nu}^j\tilde{F}_{\mu\nu}^k &= \phi \, \dfrac{1}{2}\, \epsilon^{\mu\nu\rho\sigma} \big( \partial_{_{[\mu}} G_{_{\nu]}}^j \big) \big( \partial_{_{[\rho}} G_{_{\sigma]}}^k \big)
    \, ,
\end{align}
To reconstruct the pseudo-scalar terms from the expansion of Eq.~\eqref{Tr-Q-expansion}, we need
\begin{itemize}
    \item Two insertions of $P_{\mu}$ to account for the two partial derivatives, then forming field strength tensors.
    \item One insertion of $W_1[\phi]$ to account for the pseudo-scalar field $\phi$.
    \item To account for the two gauge fields, we need $VV$ and $AA$ structures. Since $W_1[\phi]$ contains a $\gamma^5$, the combination with $AV$ structure will not contribute to the final result.
\end{itemize}
We collect the relevant classes of operators that contribute to the Wilson coefficients of these pseudo-scalar terms,
\begin{align}
    \mathcal{L}_{\rm EFT}^{\rm 1loop} &\supset i \, \tr \, \dfrac{1}{5} \int \dfrac{d^dq}{(2\pi)^d} \left[ \,  \dfrac{-1}{\slashed{q}+M} \bigg( -P_{\mu}\gamma^{\mu} -V_{\mu}[\phi]\gamma^{\mu} + A_{\mu}[\phi] \gamma^{\mu}\gamma^5 + W_1[\phi]i\gamma^5 \bigg)  \right]^5
    \nonumber \\
    &\supset \sum_N f_{_N}^{^{\pi VV}}\mathbb{O}^{(P^2V^2W_1)} + f_{_N}^{^{\pi AA}}\mathbb{O}^{(P^2A^2W_1)}
    \, , 
    \label{Goldstone: EFT expansion}
\end{align}
The evaluation of the class of operator $\mathbb{O}^{P^2V^2W_1}$ and $\mathbb{O}^{P^2A^2W_1}$ can be done very efficiently by using the One-Loop Universal Effective Action (UOLEA)~\footnote{These operators have been explicitly evaluated and are then available in the fermionic UOLEA in Ref.~\cite{Ellis:2020ivx}.}. One obtains
\begin{align}
    \mathcal{L}_{\rm EFT}^{\rm 1loop} &\supset -\dfrac{1}{8\pi^2 M} \tr \, \epsilon^{\mu\nu\rho\sigma} \bigg( W_1[P_{\mu},V_{\nu}][P_{\rho},V_{\sigma}] + \dfrac{1}{3}W_1[P_{\mu},A_{\nu}][P_{\rho},A_{\sigma}] \bigg)
    \nonumber \\
    &= \dfrac{1}{16\pi^2 M} \tr \bigg( W_1^i F^{V^j}_{\mu\nu}\tilde{F}^{V^k}_{\mu\nu} + \dfrac{1}{3} W_1^i F^{A^j}_{\mu\nu}\tilde{F}^{A^k}_{\mu\nu} \bigg)
    \, ,
    \label{uolea: pseudo-scalar-gg}
\end{align}
where we form the field strength tensors by using
\begin{align}
    \epsilon^{\mu\nu\rho\sigma}[P_{\mu},V^j_{\nu}][P_{\rho},V^k_{\sigma}] = \dfrac{i^2}{4}\epsilon^{\mu\nu\rho\sigma} \big( \partial_{[\mu}V^j_{\nu]} \big)\big( \partial_{[\rho}V^k_{\sigma]} \big)
    = -\dfrac{1}{2} F^{V^j}_{\mu\nu}\tilde{F}_{\mu\nu}^{V^k}
    \, ,
\end{align}
and similarly for the axial currents. We note that if $j\neq k$, one needs to sum over the exchange of $j,k$ indices to avoid the factor 2 problem.

\subsection{Summary and master formula}

We summarise the computations and the main outcome of section~\ref{UOLEA}. Starting with a massive fermion which bilinearly involves some, yet undetermined, vector $V_{\mu}[\phi]$, axial vector $A_{\mu}[\phi]$ and pseudo scalar $W_1[\phi]$ interactions,
\begin{align}
\mathcal{L}_{\rm UV}^{\rm fermion}\big[ \Psi, \phi \big] 
&\supset \bar{\Psi} \bigg[ i\gamma^{\mu}\partial_{\mu} - M + V_{\mu}[\phi]\gamma^{\mu} - A_{\mu}[\phi]\gamma^{\mu}\gamma^5 - W_1[\phi]i\gamma^5  \bigg]\Psi \,,
\label{LUVsum}
\end{align}
one obtains after integrating out the fermion field i.e evaluate the one-loop effective action by expanding the functional trace with CDE techniques, 
\begin{align}
\mathcal{L}_{\rm EFT}^{\rm 1loop} &= i \, \tr \sum_{n=1}^{\infty} \dfrac{1}{n} \int \dfrac{d^4q}{(2\pi)^4} \left[ \,  \dfrac{-1}{\slashed{q}+M} \bigg( -i\partial_{\mu}\gamma^{\mu} -V_{\mu}\gamma^{\mu} + A_{\mu}\gamma^{\mu}\gamma^5 + W_1 i\gamma^5 \bigg)  \right]^n \, ,
\end{align}
where in practice, the vector, axial-vector and pseudo-scalar structures are expressed as 
\begin{align}
    V_{\mu}[\phi] \equiv g_{_V}^i V_{\mu}^i[\phi^i]\, , ~
    A_{\mu}[\phi] \equiv g_{_A}^i A_{\mu}^i[\phi^i]\, , ~
    W_1[\phi] \equiv g_{_{W_1}}^i W_1^i[\phi^i] \, ,
\end{align}
with an implicit summation over the $i$ index. 
One can proceed and form the low energy effective operators and evaluate their associated Wilson coefficients which are regularised in dimensional regularisation. After regularisation, it is important to identify ambiguities of some Wilson coefficients resulting from the fact that the gauge or anomalous aspects of the symmetries have not been addressed yet.
The generic one-loop effective Lagrangian, still involving redundant operators as well as the ambiguous $\bar{a}$'s and $\bar{b}$'s coefficients, reads
\begin{tcolorbox}[height=3.8cm, valign=center, colback=white]
\begin{align}
    \mathcal{L}_{\rm EFT}^{\rm 1loop} & \supset \dfrac{1}{8\pi^2} \bar{a}_{V^iV^jA^k} \tr \bigg[  V_{\mu}^i V_{\nu}^j \tilde{F}_{\mu\nu}^{A^k} \bigg]
    + \dfrac{1}{8\pi^2} \bar{a}_{V^jA^kV^i} \tr \bigg[ V_{\mu}^j A^k_{\nu} \tilde{F}_{\mu\nu}^{V^i} \bigg] 
    + \dfrac{1}{8\pi^2} \bar{a}_{A^kV^iV^j} \tr \bigg[ A^k_{\mu} V_{\nu}^i \tilde{F}_{\mu\nu}^{V^j} \bigg] \nonumber \\
    & + \dfrac{1}{8\pi^2} \bar{b}_{A^iA^jA^k} \tr \bigg[ A^i_{\mu}A^j_{\nu} \tilde{F}_{\mu\nu}^{A^k}  \bigg] +\dfrac{1}{8\pi^2} \bar{b}_{A^jA^kA^i} \tr \bigg[ A^j_{\mu}A^k_{\nu} \tilde{F}_{\mu\nu}^{A^i}  \bigg] 
    + \dfrac{1}{8\pi^2} \bar{b}_{A^kA^iA^j} \tr \bigg[ A^k_{\mu}A^{i}_{\nu} \tilde{F}_{\mu\nu}^{A^j}  \bigg] \nonumber \\
    & + \dfrac{1}{16\pi^2 M} \tr \bigg( W_1^{i} F^{V^j}_{\mu\nu}\tilde{F}^{V^k}_{\mu\nu} + \dfrac{1}{3} W_1^{i} F^{A^j}_{\mu\nu}\tilde{F}^{A^k}_{\mu\nu} \bigg) 
    \, .
    \label{masterf}
\end{align}
\end{tcolorbox}
This master formula is generic and encapsulates all the needed computations. Indeed, at this stage, imposing the EFT to respect specific gauge invariance relations will link several of these operators together and allow to fix the ambiguities of any free Wilson coefficients in a very simple and elegant way.
Since doing so, presuppose having a concrete model in mind or set of symmetries, we now turn back to more phenomenological investigations where this master formula is applied to various models.

\section{Application to axions}\label{section: application}

In this section, we use the results obtained in section~\ref{UOLEA} to build EFT involving would-be-Goldstone bosons of spontaneously broken symmetries and Goldstone bosons of global symmetries. As a first application, we apply the master formula of Eq.~\eqref{masterf} to concretely build the intuited EFT of Eq.~\eqref{initial4op} from the toy model presented in section~\ref{toymodel}. We will then concentrate on more realistic constructions, e.g. building EFT involving the SM gauge fields and an axion or ALP. This task might precisely imply to integrate out chiral fermions which obtain their mass while the electroweak gauge symmetry is spontaneously broken, the global PQ symmetry being spontaneously and anomalously broken. We provide a simple expression adapted to SM gauge groups, and provide explicit use of it to derive axion couplings to massive gauge fields in the original 2HDM setup as proposed by Peccei and Quinn and in a more phenomenologically relevant version of it, the invisible axion DFSZ model~\cite{Dine:1981rt,Zhitnitsky:1980tq}.

\subsection{A chiral toy model}

So far, we have evaluated the operators involving three vector structures (which can also incorporate derivative couplings) and the operators involving a pseudo-scalar field which couples with two field strength tensors. We now give an example how to use the results of the previous section to derive the EFT resulting from integrating out the chiral fermion of the toy model of section~\ref{toymodel}. We remind the fermionic quadratic operator of this toy model,
\begin{align}
    \mathcal{L}_{\rm UV}^{\rm \text{toy-model}} = \bar{\Psi}\bigg[ i\partial_{\mu}\gamma^{\mu} - M + V_{\mu}\gamma^{\mu} - A_{\mu}\gamma^{\mu}\gamma^5 - W_1 i\gamma^5 \bigg] \Psi
    \label{toymodel2}
\end{align}
where the vector, axial-vector, and pseudo-scalar structures decompose as
\begin{align}
    & V_{\mu} = \bigg\{ V_{\mu},\, \bigg[ S_{\mu}-\dfrac{\partial_{\mu}\pi_S}{2v_S} \bigg] \bigg\}
    \, , ~
    A_{\mu} = \bigg\{ A_{\mu},\, \bigg[ U_{\mu}-\dfrac{\partial_{\mu}\pi_U}{2v_U} \bigg] \bigg\} 
    \, , ~
    W_{1} = M\dfrac{\pi_A}{v_A}
    \, ,    
\end{align}
and the gauge couplings are omitted for simplicity. Making use of our master formula given in Eq.~\eqref{masterf}, one can straightforwardly obtain
\begin{align}
\mathcal{L}_{\rm EFT}^{\rm 1loop}  &=
\omega_{_{VAV}} \left[ S_{\mu}-\dfrac{\partial_{\mu}\pi_{S}}{2v_{S}}\right] A_{\nu} \tilde{F}_{_V}^{\mu\nu}  +  \omega_{_{AAA}} \left[ U_{\mu}-\dfrac{\partial_{\mu}\pi_{U}}{2v_{U}}\right] A_{\nu}  \tilde{F}_{_A}^{\mu\nu}
\nonumber \\
& + \eta_{_{ASV}} \bigg[ \dfrac{\pi_A}{v_A} F_{_S,\,\mu\nu} \tilde{F}_{_V}^{\mu\nu} \bigg]  + \eta_{_{AUA}} \bigg[ \dfrac{\pi_A}{v_A} F_{_U,\,\mu\nu} \tilde{F}_{_A}^{\mu\nu} \bigg] .
\label{EFTtoymodel2}
\end{align}
At this stage, $\omega_{_{VAV}}$, $\omega_{_{AAA}}$ and $\eta_{_{ASV}}$, $\eta_{_{AUA}}$ read,
\begin{align}
\omega_{_{VAV}} = \dfrac{1}{8\pi^2}\big(1-\bar{b}\big)
\, , ~
\omega_{_{AAA}} = -\dfrac{1}{8\pi^2} \bar{a}
\, ; \quad
\eta_{_{ASV}} = \dfrac{1}{8\pi^2} 
    \, , ~
\eta_{_{AUA}} = \dfrac{1}{24\pi^2} 
    \, ,
\end{align}
with $\bar{a}$ and $\bar{b}$ the two free parameters.
As presented in section 2, we now implement the consistency between the UV model of Eq.~\eqref{toymodel2} and the associated EFT of Eq.~\eqref{EFTtoymodel2} by fixing the nature of each symmetries i.e gauge or anomalous. We identify the precise value of the parameters  $\bar{a}$ and $\bar{b}$ by requiring axial gauge invariance (leaving then the possibility that the other transformations, only, could be anomalous)
\begin{align}
    & \delta_A\bigg( \omega_{_{VAV}} \left[ S_{\mu}-\dfrac{\partial_{\mu}\pi_{S}}{2v_{S}}\right] A_{\nu} \tilde{F}_{_V}^{\mu\nu} +  \dfrac{1}{8\pi^2 }  \dfrac{\pi_A}{v_A} F_{_S,\,\mu\nu} \tilde{F}_{_V}^{\mu\nu} \bigg) = 0
    \, , ~ \text{and }
    \nonumber \\
    & \delta_A\bigg( \omega_{_{AAA}} \left[ U_{\mu}-\dfrac{\partial_{\mu}\pi_{U}}{2v_{U}}\right] A_{\nu}  \tilde{F}_{_A}^{\mu\nu} +  \dfrac{1}{24\pi^2 }  \dfrac{\pi_A}{v_A} F_{_U,\,\mu\nu} \tilde{F}_{_A}^{\mu\nu} \bigg) = 0
    \, ,
    \label{deltaA}
\end{align}
where we perform the gauge variation of the axial current, $\delta_A A_{\mu} = \partial_{\mu}\theta_A$, the would-be-Goldstone, $\delta_A \pi_A = 2 v_A \theta_A$, integrate by parts and combine the various contributions proportional to $(\partial_{\mu}\theta_A)U_{\nu}\tilde{F}_{_A}^{\mu\nu}$ and $(\partial_{\mu}\theta_A)S_{\nu}\tilde{F}_{_V}^{\mu\nu}$.
This straightforwardly leads to,  
\begin{align}
\omega_{_{VAV}}&= -\dfrac{1}{2\pi^2} ~ \Leftrightarrow ~  \bar{b} =5
    ~; \quad
\omega_{_{AAA}}= -\dfrac{1}{6\pi^2}  ~ \Leftrightarrow  ~ \bar{a} = \dfrac{4}{3}
    \, .
\end{align}
Finally, one can set to zero the artificial vector fields $S_{\mu}$ and $U_{\mu}$ and write the non-ambiguous dimension-five bosonic operators, simply as 
\begin{align}
\mathcal{L}_{\rm EFT}^{\rm 1loop} &= \dfrac{1}{2\pi^2}\dfrac{\partial^{\mu}\pi_S}{2v_S}A^{\nu}\tilde{F}_{_V,\,\mu\nu} + \dfrac{1}{6\pi^2}\dfrac{\partial^{\mu}\pi_U}{2v_U}A^{\nu}\tilde{F}_{_A,\,\mu\nu}
= -\dfrac{1}{8\pi^2}\dfrac{\pi_S}{v_S} F_{_A}^{\mu\nu}\tilde{F}_{_V,\mu\nu} - \dfrac{1}{24\pi^2}\dfrac{\pi_U}{v_U} F_{_A}^{\mu\nu}\tilde{F}_{_A,\mu\nu}
    \, .
    \label{toy model: main-result}
\end{align}
This is the one-loop contributions to the EFT Lagrangian obtained by integrating out a chiral massive fermion in our toy model. To obtain the full EFT Lagrangian, one must add the Jacobian terms given by Eq.~\eqref{UVJactot} with the one-loop terms of Eq.~\eqref{toy model: main-result},
\begin{align}
    \mathcal{L}_{\rm EFT} &=
\dfrac{1}{16\pi^{2}}\dfrac{\pi_{U}}{v_U}\left(  F_{_{V},\mu\nu} \tilde{F}^{\mu\nu}_{_V} 
+ \dfrac{1}{3} F_{_{A},\mu\nu}\tilde{F}_{_A}^{\mu\nu}\right) \, .
\end{align}
We note that integrating out the fermion in a one-dimensional representation starting from Eq.~(\ref{toymodel2}), we obtain that the $\pi_S$ derivative interaction induces EFT operators that precisely cancel the Jacobian term in Eq.~(\ref{UVJactot}) as expected starting from an abelian gauge theory, as discussed earlier, and this provides a non-trivial check for our calculation. We should also remark that the $\pi_{U}VV$ coupling entirely arises from the Jacobian term, as predicted by the Sutherland-Veltman theorem. However, the $\pi_{U}AA$ coupling does not and displays an additional factor of $1/3$ due to the one-loop contribution.

We now move on to more concrete axion models for which we will compute one-loop induced effective couplings between axions and gauge bosons, with a particular interest for those involving massive gauge fields~\footnote{These results should and will reproduce those derived, using different techniques, in Refs.~\cite{Quevillon:2019zrd,Bonnefoy:2020gyh}.}.

\subsection{Axion couplings to gauge fields}

The axion field is a relic of the spontaneous symmetry breaking of a global $U(1)_{PQ}$ symmetry.
A realistic model involving the QCD axion or an ALP, being a pseudo-scalar field $a(x)$, basically couples to fermions (of the SM or not) which have to be charged under the Global $U(1)_{PQ}$ group but also other abelian or non-abelian groups such as the one of the SM. For a massive chiral fermion, its bilinear form, after gauge symmetry breaking, generically reads
\begin{align}
\mathcal{L}_{\rm UV} = \mathcal{L}_{\rm UV}^{\rm Jac} + \bar{\Psi}\bigg[ i\partial_{\mu}\gamma^{\mu} - M + V_{\mu}\gamma^{\mu} - A_{\mu}\gamma^{\mu}\gamma^5 - W_1 i\gamma^5 \bigg]\Psi
\, ,
\label{Appl: Lagrangian-UV-generic}
\end{align}
where vector, axial-vector and pseudo-scalar structures include\footnote{Note that for convenience, we have used a different normalisation convention for the PQ charges than the one used for gauge charges.}
\begin{align}
    V_{\mu} = \big\{ g_{_V}^i V_{\mu}^i \,,\, g_{_V}^{PQ}\big(\partial_{\mu}a-V_{\mu}^{PQ}\big) \big\}
    \, , ~
    A_{\mu} = \big\{ g_{_A}^i A_{\mu}^i \, , \, g_{_A}^{PQ}\big(\partial_{\mu}a-A_{\mu}^{PQ}\big) \big\}
    \, , ~
    W_1 =  M \dfrac{\pi_{_A}^i}{v_A}
    \, ,
\end{align}
with $V_{\mu}^i,\,A_{\mu}^i$ stand for vector and axial-vector components of a generic chiral gauge field $G_{\mu}^i$, the term $\pi_{_A}^i(x)$ stands for the would-be-Goldstone boson in the case where $G_{\mu}^i$ obtains its mass from gauge spontaneous symmetry breaking. $V_{\mu}^{PQ}$ and $A_{\mu}^{PQ}$ are the fictitious auxiliary gauge fields associated to the global PQ symmetry. Writing Eq.~\eqref{Appl: Lagrangian-UV-generic} presupposes a chiral fermion reparametrisation which induces a Jacobian term, $\mathcal{L}_{\rm UV}^{\rm Jac}$. This contribution, before gauge spontaneous symmetry breaking, reads
\begin{align}
    \mathcal{L}_{\rm UV}^{\rm Jac} &= \dfrac{1}{16\pi^2f_a} \mathcal{N}_{PQ} \, a(x) F_{\mu\nu}^i \tilde{F}^{i,\,\mu\nu}
    \, ,
    \label{Appl: Jacobian-generic}
\end{align}
where the $i$-index only runs for the gauge field strength tensors. The anomaly coefficient can be generally expressed as,  
\begin{align}
    \mathcal{N}_{PQ} = \sum_{\Psi=\Psi_R,\Psi_L^{\dagger}} \tr \bigg[PQ(\Psi) \otimes G(\Psi) \otimes G(\Psi) \bigg]
    \, ,
    \label{formula: anomaly coef.}
\end{align}
with $PQ(\Psi)$ and $G(\Psi)$ the PQ and gauge charge of the chiral fermion $\Psi$. Integrating out the chiral fermion and making use of the master formula Eq.~\eqref{masterf}, one obtains 
\begin{align}
    \mathcal{L}_{\rm EFT}^{\rm 1loop} &= \omega_{_{VAV}} \bigg[\, g_{_V}^{PQ}g_{_A}^ig_{_V}^j\, \big(\partial_{\mu}a - V_{\mu}^{PQ}\big)A_{\nu}^i\tilde{F}^{V^j,\mu\nu} \bigg] + \omega_{_{AAA}}\bigg[\,g_{_A}^{PQ}g_{_A}^ig_{_A}^j\, \big(\partial_{\mu}a - A_{\mu}^{PQ} \big)A_{\nu}^i\tilde{F}^{A^j,\mu\nu} \bigg]
    \nonumber \\
    & + \dfrac{1}{8\pi^2} \big(g_{_V}^{PQ} g_{_V}^j \big) \dfrac{\pi_{_A}^i}{v_A} F_{\mu\nu}^{V^{PQ}} \tilde{F}^{V^j,\mu\nu}
    + \dfrac{1}{24\pi^2} \big(g_{_A}^{PQ} g_{_A}^j \big) \dfrac{\pi_{_A}^i}{v_A} F_{\mu\nu}^{A^{PQ}} \tilde{F}^{A^j,\mu\nu}
    \nonumber \\
    &= -\dfrac{1}{4\pi^2} \big(\, g_{_V}^{PQ}g_{_A}^ig_{_V}^j \big)\, a\,F_{\mu\nu}^{A^i}\tilde{F}^{V^j,\mu\nu} - \dfrac{1}{12\pi^2}\big(\,g_{_A}^{PQ}g_{_A}^ig_{_A}^j\big)\, a\,F_{\mu\nu}^{A^i}\tilde{F}^{A^j,\mu\nu}
    \, .
    \label{Appl: One-loop-terms}
\end{align}
In order to get the last line of the above equation, we imposed the crucial axial gauge invariance, used integration by parts and Bianchi identity, neglected the surface terms, and at the end of the computation, we removed the fictitious fields $V_{\mu}^{PQ}$ and $A_{\mu}^{PQ}$. Adding all together, we are now able to build the axion-bosonic effective Lagrangian described by 
$\mathcal{L}_{\rm EFT} = \mathcal{L}_{\rm UV}^{\rm Jac} + \mathcal{L}_{\rm EFT}^{\rm 1loop}$ where the generic formula of $\mathcal{L}_{\rm UV}^{\rm Jac}$ and $\mathcal{L}_{\rm EFT}^{\rm 1loop}$ are given by Eqs.\,\eqref{Appl: Jacobian-generic},\eqref{Appl: One-loop-terms}.

\subsubsection{SM gauge and PQ symmetries}
We now present two examples where the axion field couples with the SM gauge fields. Our first example will be the original Peccei and Quinn scenario in which the axion is the pseudo-scalar component of a Two Higgs Doublet Model (2HDM). Our second application will be to consider the so-called DFSZ axion model~\cite{Dine:1981rt,Zhitnitsky:1980tq}. To illustrate the results and properties discussed in the previous sections, we will integrate out only one generation of quarks, let us say $\big( u ~ d \big)$. This computation was performed in Ref.\cite{Quevillon:2019zrd} by Feynman diagram technique, accompanied by Pauli-Villars regularisation. We will recover some of its results by using the functional method for one-loop matching.

We begin with the Jacobian terms which induce tree-level axion couplings to the SM gauge fields,
\begin{align}
    \mathcal{L}_{\rm UV}^{\rm Jac} &= \dfrac{1}{16\pi^2f_a}\, \bigg( g_s^2\mathcal{N}_C \,  a G_{\mu\nu}\tilde{G}^{\mu\nu} + g^2\mathcal{N}_L\, a W_{\mu\nu}^i\tilde{W}^{i,\mu\nu} + g'^2\mathcal{N}_Y\, a B_{\mu\nu}\tilde{B}^{\mu\nu} \bigg)
    \, ,
    \label{jacobian: PQ-SM}
\end{align}
with the anomaly coefficients $\mathcal{N}_i$ computable as follows,  
\begin{align}
    & \mathcal{N}_C = \sum_{\Psi=q_L^{\dagger},u_R,d_R} C_{_{SU(3)_c}}(\Psi)\, d_{_{SU(2)_L}}(\Psi) \, PQ(\Psi)
    \, ,
    \nonumber \\
    & \mathcal{N}_L = \sum_{\Psi=q_L^{\dagger};\,l_L^{\dagger}} d_{_{SU(3)_c}}(\Psi)\, C_{_{SU(2)_L}}(\Psi) \, PQ(\Psi)
    \, ,
    \nonumber \\
    & \mathcal{N}_Y = \sum_{\Psi=q_L^{\dagger},u_R,d_R;\,l_L^{\dagger},e_R} d_{_{SU(3)_c}}(\Psi)\, d_{_{SU(2)_L}}(\Psi) C_{_{U(1)_Y}}(\Psi) \, PQ(\Psi)
    \, ,
    \label{anomaly coeff: PQ-SM}
\end{align}
where we closely followed the conventions and notations of Ref.\cite{Quevillon:2019zrd} with $d_{_{SU(3)_c}}(\Psi)$, $d_{_{SU(2)_L}}(\Psi)$ and $C_{_{SU(3)_c}}(\Psi)$, $C_{_{SU(2)_L}}(\Psi)$ are respectively the $SU(3)_c$ and $SU(2)_L$ dimensions and quadratic Casimir invariant of the representation carried by the chiral fermion field $\Psi$. Besides, $PQ(\Psi)$ is the PQ charge of the fermion $\Psi$ which is model-dependent. We will come back to these PQ charges when discussing a peculiar axion model.

The one-loop effective Lagrangian resulting from integrating out the SM chiral fermion is  
\begin{align}
    \mathcal{L}_{\rm EFT}^{\rm 1loop} \supset \sum_f  \dfrac{-1}{4\pi^2}  \bigg[ & \big( g_{_V}^{PQ}g_{_A}^Zg_{_V}^Z \big)^f \, \bigg( a\, F_{\mu\nu}^{A^Z} \tilde{F}^{V^Z,\mu\nu} \bigg) + \dfrac{1}{3}\big( g_{_A}^{PQ}g_{_A}^Zg_{_A}^Z \big)^f \, \bigg( a\, F_{\mu\nu}^{A^Z}\tilde{F}^{A^Z,\mu\nu} \bigg)
    \nonumber \\
    + & \big( g_{_V}^{PQ}g_{_A}^Wg_{_V}^W \big)^f \, \bigg( a\, F_{\mu\nu}^{A^W} \tilde{F}^{V^W,\mu\nu} \bigg) + \dfrac{1}{3}\big( g_{_A}^{PQ}g_{_A}^Wg_{_A}^W \big)^f \, \bigg( a\, F_{\mu\nu}^{A^W} \tilde{F}^{A^W,\mu\nu} \bigg)
    \nonumber \\
    + & \big( g_{_V}^{PQ}g_{_A}^Zg_{_V}^{\gamma} \big)^f \, \bigg( a\, F_{\mu\nu}^{A^Z} \tilde{F}^{V^{\gamma},\mu\nu} \bigg) \bigg]
    \, ,
    \label{loopEFT: PQ-SM-gauge}
\end{align}
where $g_{_V}^{PQ},\,g_{_A}^{PQ}$ are axion-fermion-fermion couplings written in terms of Dirac bilinear form. A summary of the gauge charges of SM fermions can be found in Table\,\ref{tab:SM-gauge-fermion}.
\begin{table}[h!]
    \centering
    \begin{tabular}{c|cccc}
        $i$ & g & W & $\gamma$ & Z 
        \\ \hline
        \T\B $\big(g_{_{V}}^{\,i}\big)^f$ & $g_s T_a^f$ & $\dfrac{g}{\sqrt{2}}T_3^f$ & $eQ^f$ &  $\dfrac{g}{2\cos\theta_w} (T_3^f - 2\sin^2\theta_w Q^f)$ 
        \\ \hline
        \T\B $\big(g_{_{A}}^{\,i}\big)^f$ & 0 & $\dfrac{g}{\sqrt{2}}T_3^f$ & 0 & $\dfrac{g}{2\cos\theta_w}T_3^f$ 
    \end{tabular}
    \caption{SM fermion couplings to the SM gauge fields, where $T_a^f,\,T_3^f,\,Q^f,\,\theta_w$ are respectively the $SU(3)_C$ generators,  the eigenvalue of the isospin operator, the electromagnetic charge and the weak mixing angle.  
    }
    \label{tab:SM-gauge-fermion}
\end{table}
\\
The only thing that remains to be determined in Eqs.~\eqref{jacobian: PQ-SM},\,\eqref{anomaly coeff: PQ-SM},\,\eqref{loopEFT: PQ-SM-gauge} are the fermions PQ charge, that we discuss now for several axion models.

\subsubsection{PQ axion model}
We first consider the original PQ scenario where the QCD axion is identified as the orthogonal state of the would-be-Goldstone of the Z boson in a 2HDM model (see Refs.~\cite{Peccei:1977hh,Peccei:1977ur}). 
The starting point is a fermion-Higgs Yukawa interaction, that we assume of type II, which can be written as
\begin{align}
    \mathcal{L}_{\rm Yukawa}^{\rm 2HDM} &= -\bigg[  Y_u\bar{u}_R\, \Phi_1\, q_L + Y_d\bar{d}_R\,\Phi_2^{\dagger}\,q_L \bigg] - Y_e\bar{e}_R\,\Phi_2^{\dagger}\,l_L + \text{h.c.}
    \, . 
    \label{Yukawa: 2HDM}
\end{align}
The two complex scalar fields can be written as
\begin{align}
    \Phi_1 = \dfrac{1}{\sqrt{2}} e^{i\frac{\eta_1}{v_1}}
    \begin{pmatrix}
    0 \\v_1
    \end{pmatrix}
    \, , ~
    \Phi_2 = \dfrac{1}{\sqrt{2}} e^{i\frac{\eta_2}{v_2}}
    \begin{pmatrix}
    0 \\v_2
    \end{pmatrix}
    \, ,
    \label{2HDM: Higgs-doublets}
\end{align}
where $\eta_1,\,\eta_2$ are Goldstone bosons of the scalar fields $\Phi_1$ and $\Phi_2$. The vacuum expectation value of the scalar fields, $v_1$ and $v_2$ are related by $v_1^2 + v_2^2 \equiv v^2 \simeq \big(246\,\text{GeV}\big)^2$, and one usually introduces the $\beta$ angle such that $v_1 = v\sin\beta$, $v_2=v\cos\beta$ and $v_2 / v_1 = \big(1 / \tan\beta\big) \equiv x$. The next step is to identify the would-be-Goldstone boson (that generates the mass of the Z-boson) from its orthogonal state, defining then the axion. One has the following relations
\begin{align}
    \begin{pmatrix}
    G^0 \\ a\, 
    \end{pmatrix}
    = 
    \begin{pmatrix}
    \cos\beta & \sin\beta \\
    -\sin\beta & \cos\beta
    \end{pmatrix}
    \begin{pmatrix}
    \eta_2 \\ 
    \eta_1
    \end{pmatrix}
    \, .
    \label{diagonal Goldstone-axion}
\end{align}
The Higgs doublets can be re-written as
\begin{align}
    \Phi_1 = \dfrac{1}{\sqrt{2}}e^{i\frac{G^0}{v_1}}e^{i\,x\frac{a\, }{v}}
    \begin{pmatrix}
    0 \\ v_1
    \end{pmatrix}
    \, , ~
    \Phi_2 = \dfrac{1}{\sqrt{2}}e^{i\frac{G^0}{v_2}}e^{i\,\big(-\frac{1}{x}\big)\frac{a\, }{v}}
    \begin{pmatrix}
    0 \\ v_2
    \end{pmatrix}
    \, ,
    \label{2HDM: Higgs-doublet-axion}
\end{align}
where $G^0$ is PQ neutral and the Higgs doublets carry the following PQ charge, $PQ(\Phi_1)=x$ and $PQ(\Phi_2)=-1/x$. In order to identify the PQ axion model with Eq.~\eqref{Appl: Lagrangian-UV-generic}, we first make the Yukawa Lagrangian becomes PQ-invariant by performing the chiral rotation,
\begin{align}
    \Psi \rightarrow e^{iPQ(\Psi)\frac{a}{v}} \Psi
    \, .
\end{align}
The PQ charges for one generation of quarks $\big(u ~ d \big)$ are assigned, such as
\begin{align}
    PQ\big(q_L;u_R,d_R\big) = \bigg(\alpha;\alpha+x,\alpha+\dfrac{1}{x}\bigg)
    \, .
    \label{PQ charges: LR-2HDM}
\end{align}
$\alpha$ is a free parameter that corresponds to the conservation of the baryon number\footnote{For a general setup including also the lepton sector see Refs.\cite{Quevillon:2019zrd,Quevillon:2020hmx}.}. The chiral rotation leads to the derivative coupling of axion with SM fermions as defined in Eq.~\eqref{Appl: Lagrangian-UV-generic} and the axion couplings to fermions read
\begin{align}
   & \big( g_{_V}^{PQ} \big)^u = -\dfrac{1}{2v}(2\alpha + x) 
   \, , ~ 
   \big( g_{_A}^{PQ} \big)^u = \dfrac{1}{2v} x
   \, ; \quad
   \big( g_{_V}^{PQ} \big)^d = -\dfrac{1}{2v}\bigg(2\alpha + \dfrac{1}{x}\bigg) 
   \, , ~ 
   \big( g_{_A}^{PQ} \big)^d = \dfrac{1}{2v}\bigg( \dfrac{1}{x} \bigg)
   \, .
   \label{PQ charges: VA-2HDM}
\end{align}
Plugging Eq.~\eqref{PQ charges: LR-2HDM} into Eq.~\eqref{jacobian: PQ-SM} and rotating the electroweak gauge fields from their interaction basis to their physical mass basis using $W_{\mu}^3=c_wZ_{\mu}+s_wA_{\mu},\, B_{\mu}=-s_wZ_{\mu}+c_wA_{\mu}$ along with $e=gs_w=g'c_w$, one obtains the following Lagrangian for the Jacobian contribution
\begin{align}
    \mathcal{L}_{\rm Jac}^{\{u,d\}} =
    \dfrac{1}{16\pi^2v} \bigg( & \dfrac{g_s^2}{2}\bigg[x+\dfrac{1}{x}\bigg] a\,G_{\mu\nu}^a \tilde{G}^{a,\mu\nu} + e^2N_c \bigg[ \dfrac{4}{9}x + \dfrac{1}{9x}  \bigg] a\,F_{\mu\nu}\tilde{F}^{\mu\nu} 
    \nonumber \\
    & -\big[ g^2N_c \, \alpha \big]a\, W_{\mu\nu}^+\tilde{W}^{-,\mu\nu}
    - \dfrac{2e^2}{c_ws_w}N_c\bigg[ \dfrac{1}{2}\alpha + s_w^2\bigg( \dfrac{4}{9}x + \dfrac{1}{9x} \bigg) \bigg]a\, Z_{\mu\nu}\tilde{F}^{\mu\nu}
    \nonumber \\
    & + \dfrac{e^2}{c_w^2s_w^2}N_c\bigg[ -(1-2s_w^2)\dfrac{\alpha}{2} + s_w^4\bigg( \dfrac{4}{9}x + \dfrac{1}{9x} \bigg) \bigg]a\, Z_{\mu\nu}\tilde{Z}^{\mu\nu} \bigg) \, ,
\end{align}
where $N_c = 3$. Plugging Eq.~\eqref{PQ charges: VA-2HDM} into Eq.~\eqref{loopEFT: PQ-SM-gauge} and performing the same electroweak rotation lead to the following one-loop effective Lagrangian,
\begin{align}
    \mathcal{L}_{\rm EFT}^{\rm 1loop-\{u,d\}} &= \dfrac{1}{16\pi^2 v} \bigg( g^2N_c \bigg[ \alpha + \dfrac{1}{6}\bigg(x+\dfrac{1}{x}\bigg) \bigg]a\, W_{\mu\nu}^+\tilde{W}^{-,\mu\nu}
    + \dfrac{e^2}{c_ws_w}N_c\bigg[ \alpha + \bigg(\dfrac{1}{3}x+\dfrac{1}{6x}\bigg) \bigg]a\, Z_{\mu\nu}\tilde{F}^{\mu\nu}
    \nonumber \\
    &\qquad\qquad~\, + \dfrac{e^2}{c_w^2s_w^2}N_c\bigg[ (1-2s_w^2)\dfrac{\alpha}{2} + \dfrac{1}{12}\bigg(x+\dfrac{1}{x}\bigg) - s_w^2\bigg(\dfrac{1}{3}x+\dfrac{1}{6x}\bigg) \bigg]a\, Z_{\mu\nu}\tilde{Z}^{\mu\nu} \bigg) \, .
\end{align}
The effective axion-bosonic Lagrangian is obtained by adding $\mathcal{L}_{\rm Jac}^{\{u,d\}}$ and $\mathcal{L}_{\rm EFT}^{\rm 1loop-\{u,d\}}$ and gives the compact result,
\begin{align}
\mathcal{L}_{\rm EFT}^{\rm a-bosonic} &= \dfrac{1}{16\pi^2v} \bigg(  \dfrac{g_s^2}{2}\bigg[x+\dfrac{1}{x}\bigg] a\,G_{\mu\nu}^a \tilde{G}^{a,\mu\nu} + e^2N_c \bigg[ \dfrac{4}{9}x + \dfrac{1}{9x}  \bigg] a\,F_{\mu\nu}\tilde{F}^{\mu\nu}  
\nonumber \\
& \, + g^2N_c \,\dfrac{1}{6}\bigg[x+\dfrac{1}{x}\bigg] a\, W_{\mu\nu}^+\tilde{W}^{-,\mu\nu} 
+ \dfrac{e^2}{c_w s_w}N_c\bigg[ \bigg(\dfrac{1}{3}x + \dfrac{1}{6x}\bigg) - 2s_w^2\bigg(\dfrac{4}{9}x+\dfrac{1}{9x} \bigg)  \bigg]a\, Z_{\mu\nu}\tilde{F}^{\mu\nu}
\nonumber \\
& \, + \dfrac{e^2}{c_w^2s_w^2}N_c\bigg[ \dfrac{1}{12}\bigg(x+\dfrac{1}{x}\bigg) - s_w^2\bigg(\dfrac{1}{3}x+\dfrac{1}{6x}\bigg) + s_w^4\bigg( \dfrac{4}{9}x + \dfrac{1}{9x} \bigg) \bigg] a\, Z_{\mu\nu}\tilde{Z}^{\mu\nu} \bigg) \, .
\label{a-bosonicEFT: 2HDM}
\end{align}

\subsubsection{DFSZ axion model}

Concerning the case of the more realistic axion DFSZ model~\cite{Dine:1981rt,Zhitnitsky:1980tq}, the Yukawa couplings are the same as in the 2HDM model, but now the scalar potential is modified. Typically, the 2HDM model is extended by a gauge-singlet complex scalar field $\phi$, with the scalar potential
\begin{align}
    V_{\rm DFSZ} &= V_{\rm 2HDM} + V_{\phi 2HDM} + V_{\phi PQ} + V_{\phi}
    \, ,
\end{align}
where we have
\begin{align}
    \begin{cases}
    V_{\phi 2HDM} = a_1\big(\phi^{\dagger}\phi \big)\big(\Phi^{\dagger}_1\Phi_1 \big) + a_2\big(\phi^{\dagger}\phi \big)\big(\Phi^{\dagger}_2\Phi_2 \big) \, ,
    \\
    V_{\phi PQ} = \lambda_{12}\big(\phi^{\dagger}\phi \big)\Phi_1^{\dagger}\Phi_2 + \text{h.c.} \, ,
    \\
    V_{\phi} = \mu^2\big(\phi^{\dagger}\phi \big) + \lambda\big(\phi^{\dagger}\phi \big)^2 \, .
    \end{cases}
\end{align}
Similarly to $\Phi_i$ of Eq.~\eqref{2HDM: Higgs-doublet-axion}, one can also write the new complex scalar field $\phi$ as
\begin{align}
    \phi = \dfrac{1}{\sqrt{2}}e^{i\frac{\eta_a}{f_a}}
    \begin{pmatrix}
    0 \\ f_a
    \end{pmatrix} \, .
\end{align}
In summary, for the DFSZ axion model, one obtains the PQ-charges and the breaking-scale of the PQ-symmetry by rescaling their values in the axion PQ model, simply as follows,
\begin{align}
    x \rightarrow \dfrac{2x^2}{x^2+1}
    \, , ~
    \dfrac{1}{x} \rightarrow \dfrac{2}{x^2+1}
    \, , ~
    v \rightarrow f_a
    \, .
    \label{DFSZ rescale}
\end{align}
The effective DFSZ axion-bosonic Lagrangian, obtained by adding $\mathcal{L}_{\rm Jac}^{\{u,d\}}$ and $\mathcal{L}_{\rm EFT}^{\rm 1loop-\{u,d\}}$, is given by
\begin{align}
\mathcal{L}_{\rm EFT}^{\rm a-bosonic}
= \dfrac{1}{16\pi^2f_a} \bigg( &
g_s^2\, a\,G_{\mu\nu}^a \tilde{G}^{a,\mu\nu} + e^2N_c \dfrac{8x^2+2}{\,9(x^2+1)} a\,F_{\mu\nu}\tilde{F}^{\mu\nu}  
\nonumber \\
& + \dfrac{g^2N_c}{3} a\, W_{\mu\nu}^+\tilde{W}^{-,\mu\nu} 
+ \dfrac{e^2}{c_w s_w}N_c \dfrac{3+6x^2-4s_w^2(4x^2+1)}{9(x^2+1)} a\, Z_{\mu\nu}\tilde{F}^{\mu\nu}
\nonumber \\
& + \dfrac{e^2}{c_w^2s_w^2}N_c\bigg[ \dfrac{1}{6} - s_w^2\dfrac{2x^2+1}{3(x^2+1)} + s_w^4 \dfrac{8x^2+2}{9(x^2+1)} \bigg] a\, Z_{\mu\nu}\tilde{Z}^{\mu\nu} \bigg) \, .
\label{a-bosonicEFT: DFSZ}
\end{align}

These results do agree with those derived in Refs.~\cite{Quevillon:2019zrd}, using the more traditional approach of Feynman diagram computations.

It is certainly a good moment to pause and appreciate the difference in strategy with this last reference.
The main and obvious distinction is that in this work, we favored the path integral method to evaluate one-loop processes. However, we believe that another elegant and insightful feature of this axionic EFT derivation is due to the direct and consistent way of dealing with gauge and anomalous symmetries. Indeed, one needs not to use the anomalous Ward-identities to alleviate ambiguities inherent to anomalies in QFTs. Equivalently, one can use the interplay between higher-dimensional operators involving the axion and the would-be-Goldstone bosons in order to consistently and easily derive axion EFTs. This offers a neat method to also explore other sectors of axion EFTs.

\section{Conclusion\label{Ccl}}

In this work, we have considered the task of building EFTs by integrating out fermions charged under both local and global symmetries. These symmetries can be spontaneously broken, and the global ones might also be anomalously broken. This setting is typically that encountered in axion models, where a new global but anomalous symmetry, $U(1)_{PQ}$, is spontaneously broken, so as to generate a Goldstone boson, the axion, coupled to gluons.

The main novelties of our approach are twofold. First, the heavy fermion to be integrated out is allowed to have chiral charges for both the local and global symmetries. The analysis is then much more intricate because of the presence of anomalies in various currents, and because the fermion can only have a mass when all the chiral symmetries are spontaneously broken. Second, we perform our analysis in a functional approach, by systematically building EFTs using an inverse mass expansion, that is, identifying leading operators and calculating their Wilson coefficients with the help of Covariant Derivative Expansion. Our calculations are adaptable to various UV models and allow us to correctly treat QFT anomalies.

In more details, our main results are the following:
\begin{itemize}
\item It exists many motivations for introducing Goldstone bosons of global symmetries using a polar representation. Once this choice has been made, we have identified an appropriate parametrisation of the fermionic part of the UV Lagrangian. Essentially, with the purpose of an inverse mass expansion, if one wants to perform an exact computation without truncating the initial UV theory, it is desirable to write the fermion mass term as an invariant quantity under the various symmetries, even for a chiral fermion. This requires some fermion field redefinitions. Only then one can clearly identify the fermion bilinear operator to be inverted.

\item Usually, Ward identities are used to enforce the desired gauge symmetries. When dealing with anomalous quantities, these constraints are crucial to remove the ambiguities that creep in through the regularisation process. But in our functional approach, this cannot be immediately implemented because the leading operators in the EFT end up being automatically gauge invariant. The only way forward is to perturb the theory to upset this automatic gauge invariance. This is done with the help of background fields, in a way very similar as in Ref.~\cite{Bonnefoy:2020gyh}. Then, the necessary Ward identity constraints can be recovered thanks to EFT operators involving the would-be-Goldstone bosons of the exact gauge symmetries.

\item The parametrisation of the fermion bilinear operator involves derivative interactions with scalar and pseudoscalar fields. To our knowledge, a precise description of how to perform the calculation of the determinant of such operators has never been presented. It should be noted that in that calculation, regularisation is necessary. For that, we adopt dimensional regularisation and follow the 't Hooft-Veltman prescription. We show that the two-parameter ambiguities, well known in the context of triangle Feynman diagrams, can be recovered. Those are crucial to allow one to enforce all the gauge constraints in a consistent way.

\item We recover in the functional context the results of Refs.~\cite{Quevillon:2019zrd, Quevillon:2020hmx, Bonnefoy:2020gyh}, that is, that the derivative coupling of the Goldstone boson $\pi$ to the fermions, $\bar{\Psi}(\partial_{\mu}\pi\,\gamma^{\mu}\gamma^{5})\Psi$ and $\bar{\Psi}(\partial_{\mu}\pi\,\gamma^{\mu})\Psi$, do not necessarily vanish in the infinite mass limit. They do contribute to the leading EFT operator $\pi VA$, $\pi AA$, but not $\pi VV$. In other words, this last coupling satisfies the Sutherland-Veltman theorem, and is fully driven by the anomaly, but not the other two. 
\end{itemize}

In this paper we have presented how to deal with scenarios combining both spontaneous and anomalous symmetry breaking. When building an EFT by integrating out chiral fermions charged under those various symmetries it is legitimate to keep local partial derivative interactions instead of traditional pseudo-scalar ones, but this has a cost. Now the anomaly is spread into several contributions which have to be recombined with high care when evaluating the S-matrix (see also Refs.~\cite{Quevillon:2019zrd, Quevillon:2020hmx, Bonnefoy:2020gyh}). We have integrated these peculiar fermions in the elegant and minimal functional approach and showed how to remove the ambiguities one has to face to evaluate the functional trace in dimensional regularisation. Inevitably, this corresponds to implement the anomalous Ward identities in a consistent way within the path integral formalism. We did so by introducing fictitious vector fields associated to the global symmetries so one can cure potential ambiguities undermining the theory while enforcing gauge invariance. More generally, this work shows a possible, neat and systematic path to follow to consistently build an entire EFT involving anomalous symmetries. It should also be very useful to derive other EFT higher dimensional operators. All in all, this procedure allowed us to compute in a transparent and in a very generic way the Wilson coefficients of higher dimensional operators involving Goldstone bosons, this is encapsulated in the master formula Eq.~\eqref{masterf}. Furthermore, we showed how to apply this master formula to the case of SM gauge interactions. Ultimately, we applied these results to the axion Goldstone boson (in the general sense i.e being the QCD axion or simply an ALP). We obtained in a closed form the higher dimensional operators involving the axion and SM gauge fields and collected them so that one can recover the non-intuitive physical coupling between axions and massive SM gauge fields which have been recently derived by some of us in Ref.~\cite{Quevillon:2019zrd}. The phenomenological relevance of these couplings are of particular interest for collider ALPs searches but also their imprints in the early universe.

\subsubsection*{Acknowledgements} 
We thank Quentin Bonnefoy and Christophe Grojean for helpful discussions. J.Q. and P.N.H.V. have benefited from fruitful discussions and collaborations with Sebastian Ellis, Tevong You and Zhengkang Zhang on several past works on functional matching. This work is supported by the IN2P3 Master project ``Axions from Particle Physics to Cosmology''. J.Q.'s work is also supported by the  IN2P3 Master project Théorie-BSMGA and the IN2P3 Master project UCMN. J.Q. acknowledges support by Institut Pascal at Université Paris-Saclay during the Paris-Saclay Astroparticle Symposium 2021, with the support of the P2IO Laboratory of Excellence (program “Investissements d’avenir” ANR-11-IDEX-0003-01 Paris-Saclay and ANR-10-LABX-0038), the P2I axis of the Graduate School Physics of Université Paris-Saclay, as well as IJCLab, CEA, IPhT, APPEC and EuCAPT.

\appendix

\section{Master integrals}\label{Appendix:master_integrals}
In this appendix, we discuss the master integrals and tabulate some of
them that are useful in practice. In this paper our results are written in terms of master integrals $\,\mathcal{I}$, defined by

\begin{align}
            \int \dfrac{d^dq}{(2\pi)^d} \dfrac{q^{\mu_1}\cdots q^{\mu_{2n_c}}}{(q^2-M_i^2)^{n_i}(q^2-M_j^2)^{n_j}\cdots }
            = g^{\mu_1 \cdots \mu_{2n_c}} \mathcal{I}[q^{2n_c}]^{n_i n_j \cdots}_{i j \cdots }
\end{align}
with;
\begin{equation}
\mathcal{I}[q^{2n_c}]_i^{n_i} = \frac{i}{16\pi^2} \bigl(-M_i^2\bigr)^{2+n_c-n_i}
\frac{1}{2^{n_c}(n_i-1)!} \frac{\Gamma(\frac{\epsilon}{2}-2-n_c +n_i)}{\Gamma(\frac{\epsilon}{2})} \Bigl(\frac{2}{\epsilon} -\gamma +\log 4\pi-\log\frac{M_i^2}{\mu^2}\Bigr) \,,
\end{equation}
where $d=4-\epsilon$ is the space-time dimension, and $\mu$ is the renormalisation scale. In the $\overline{\rm MS}$ scheme, we replace, $\bigg(\dfrac{2}{\epsilon} -\gamma \, + \, \log 4\pi -\log\dfrac{M_i^2}{\mu^2}\bigg)$ by $\bigg(-\log\dfrac{M_i^2}{\mu^2}\bigg)$ in the final result. We factor out the common prefactor, $\mathcal{I}=\frac{i}{16\pi^2}\tilde{\mathcal{I}}$ and present a table of $\tilde{\mathcal{I}}[q^{2n_c}]_i^{n_i}$ for various $n_c$ and $n_i$, needed in our computations, in Table~\ref{tab:MIheavy}.

\begin{table}[htbp!]
\centering
\begin{tabular}{|c|ccc|}
\hline
$\tilde{\mathcal{I}}[q^{2n_c}]_i^{n_i}$ & $n_c=0$ & $n_c=1$ & $n_c=2$  \\
\hline
$n_i=1$ \T
& $M_i^2 \bigl(1-\logm{M_i^2}\bigr)$ 
& $\frac{M_i^4}{4} \bigl(\frac{3}{2} -\logm{M_i^2}\bigr)$ 
& $\frac{M_i^6}{24} \bigl(\frac{11}{6} -\logm{M_i^2}\bigr)$ \\
$n_i=2$ 
& $-\logm{M_i^2}$ 
& $\frac{M_i^2}{2} \bigl(1 -\logm{M_i^2}\bigr)$ 
& $\frac{M_i^4}{8} \bigl(\frac{3}{2} -\logm{M_i^2}\bigr)$ \\
$n_i=3$ 
& $-\frac{1}{2M_i^2}$ 
& $-\frac{1}{4}\logm{M_i^2}$ 
& $\frac{M_i^2}{8} \bigl(1 -\logm{M_i^2}\bigr)$ \\
$n_i=4$ 
& $\frac{1}{6M_i^4}$ 
& $-\frac{1}{12M_i^2}$
& $-\frac{1}{24}\logm{M_i^2}$\\
$n_i=5$
& $-\frac{1}{12M^6}$
& $\frac{1}{48M^4}$
& $-\frac{1}{96M^2}$ \B \\
\hline
\end{tabular}
\caption{Commonly-used master integrals with degenerate heavy particle masses. $\tilde{\mathcal{I}}=\mathcal{I}/\frac{i}{16\pi^2}$ and the $\frac{2}{\epsilon} -\gamma +\log 4\pi$ contributions are dropped.}
\label{tab:MIheavy}
\end{table}

\pagebreak

\bibliographystyle{JHEP}
\bibliography{biblio}
\end{document}